\documentclass[runningheads]{llncs}
\usepackage{tikz,amsmath, amssymb,bm,color}
\usetikzlibrary{positioning, calc,chains,fit,shapes,decorations.pathmorphing}
\usepackage{comment}
\usepackage{tikz}
\usepackage{graphics}
\usepackage{longtable}
\usepackage{mdframed}
\usepackage{caption}
\usepackage{subcaption}
\usepackage{slashbox}
\usepackage{url}
\usepackage{framed}
\usepackage{array}
\usepackage{tabu}
\usepackage{lscape}
\usepackage{multirow}
\usepackage{ulem}
\usepackage{multicol}
\usepackage{placeins}
\usepackage{cite}
\newtheorem{observation}{Observation}
\newtheorem{construction}{Construction}

\newcommand{\pname}[1]{\textnormal{\textsc{#1}}}

\newcommand{\HFC}{\pname{${H}$-free Contraction}}
\newcommand{\PC}{\pname{$\Pi$-Contraction}}

\newcommand{\VC}{\pname{Vertex Cover}}
\newcommand{\MSOO}{MSO\textsubscript{1}}

\newcommand{\DS}{\textsc{Dominating Set}}
\newcommand{\HC}{\textsc{$H$-Contraction}}
\newcommand{\DomaticNum}{\textsc{Domatic Number}}
\newcommand{\NPC}{NP-complete}
\newcommand{\XFC}[1]{\pname{$#1$-free Contraction}}
\newcommand{\SURP}{cost$(\mathcal{P})$}
\newcommand{\SURPD}{cost$(\mathcal{P}')$}
\newcommand{\SURPC}{cost$(\mathcal{\hat{P}})$}

\newcommand{\HED}{\pname{$H$-free Edge Deletion}}
\newcommand{\HEC}{\pname{$H$-free Edge Completion}}
\newcommand{\HEE}{\pname{$H$-free Edge Editing}}
\newcommand{\HVD}{\pname{$H$-free Vertex Deletion}}

\usetikzlibrary{fit}
\usetikzlibrary{
  graphs,
  graphs.standard
}

\usepackage[shortlabels]{enumitem}
\usepackage{comment}
\setlistdepth{9}

\newlist{myEnumerate}{enumerate}{9}
\setlist[myEnumerate,1]{label=(\arabic*)}
\setlist[myEnumerate,2]{label=(\Roman*)}
\setlist[myEnumerate,3]{label=(\Alph*)}
\setlist[myEnumerate,4]{label=(\roman*)}
\setlist[myEnumerate,5]{label=(\alph*)}
\setlist[myEnumerate,6]{label=(\arabic*)}
\setlist[myEnumerate,7]{label=(\Roman*)}
\setlist[myEnumerate,8]{label=(\Alph*)}
\setlist[myEnumerate,9]{label=(\roman*)}

\begin{document}
\title{Contracting edges to destroy a pattern: A complexity study\thanks{This work is partly sponsored by SERB(India) grants ``Complexity dichotomies for graph modification problems'' (SRG/2019/002276), and  ``Algorithmic study on hereditary graph properties'' (MTR/2022/000692), and a public grant overseen by the French National Research Agency as part of the “Investissements d’Avenir” through the IMobS3 Laboratory of Excellence (ANR-10-LABX-0016), and the IDEX-ISITE initiative CAP 20-25 (ANR-16-IDEX-0001). We also acknowledge support of the ANR project GRALMECO (ANR-21-CE48-0004).}}
%
%
\author{Dipayan Chakraborty\inst{1,}\inst{2} \and
R. B. Sandeep\inst{2}}
\authorrunning{D. Chakraborty and R. B. Sandeep}
%
\institute{LIMOS, Université Clermont Auvergne, France, \email{dipayan.chakraborty@uca.fr}\and
Department of Mathematics and Applied Mathematics, University of Johannesburg, South Africa\and
Department of Computer Science and Engineering, Indian Institute of Technology Dharwad, India, \email{sandeeprb@iitdh.ac.in}}
\maketitle              
\begin{abstract}    
Given a graph $G$ and an integer $k$, the objective of the 
\PC\ problem is to check whether there exists at most $k$ 
edges in $G$ such that contracting them in $G$ results 
in a graph satisfying the property $\Pi$. We investigate the 
problem where $\Pi$ is `$H$-free' 
(without any induced copies of $H$). 
It is trivial that \XFC{H} is polynomial-time solvable 
if $H$ is a complete graph of at most two vertices. 
We prove that, in all other cases, the problem is 
NP-complete. We then investigate the fixed-parameter 
tractability of these problems. We prove that whenever 
$H$ is a tree, except for seven trees, \XFC{H} is W[2]-hard. 
This result along with the known results leaves behind only three unknown cases among trees.

\end{abstract}
\section{Introduction}

Let $\Pi$ be any graph property. Given a graph $G$ and an integer $k$, the objective of the \PC\ problem is to check whether $G$ contains 
at most $k$ edges so that contracting them results in a graph with
property $\Pi$. 
This is a vertex partitioning problem in disguise: Find whether 
there is a partition $\mathcal{P}$ of the vertices of $G$ such that
each set in $\mathcal{P}$ induces a connected subgraph of $G$, 
$G/\mathcal{P}$ (the graph obtained by contracting each set in $\mathcal{P}$ into a vertex) has property $\Pi$, and $n-|\mathcal{P}|\leq k$.
These problems, for various graph properties $\Pi$,
have been studied for the last four decades. Asano and Hirata~\cite{asano1983edge} proved that the problem is \NPC\ if 
$\Pi$ is any of the following classes - planar, series-parallel, outerplanar, chordal. When $\Pi$ is a singleton set $\{H\}$, then
the problem is known as $\HC$. Brouwer and Veldman~\cite{brouwer1987contractibility} proved that \HC\ is polynomial-time solvable if $H$ is a star, and \NPC\ if $H$ is a
connected triangle-free graph other than a star graph. 
Belmonte, Heggernes, and van 't Hof~\cite{DBLP:journals/dam/BelmonteHH12} proved that it is polynomial-time solvable when $H$ is a split graph.
Golovach, Kaminski, Paulusma, and Thilikos~\cite{DBLP:journals/tcs/GolovachKPT13} studied the problem when $\Pi$ is `minimum degree at least $d$'  and proved that the problem is \NPC\ even for $d=14$ and W[1]-hard when parameterized by $k$. Heggernes, van 't Hof, Lokshtanov, and Paul~\cite{DBLP:journals/siamdm/HeggernesHLP13} proved that the problem is fixed parameter tractable when $\Pi$ is the class of bipartite graphs. Guillemot and Marx~\cite{DBLP:journals/ipl/GuillemotM13} obtained a faster FPT algorithm for the problem. 
Cai, Guo~\cite{DBLP:conf/iwpec/CaiG13}, and Lokshtanov, Misra, and Saurabh~\cite{DBLP:conf/iwpec/LokshtanovMS13} proved that the 
problem is W[2]-hard when $\Pi$ is the class of chordal graphs.
Garey and Johnson~\cite{DBLP:books/fm/GareyJ79} mentioned that, given two graphs $G$ and $H$, the problem of checking whether $H$ can be 
obtained from $G$ by edge contractions is \NPC. 
Edge contraction has applications in Graph minor theory (see \cite{lovasz2006graph}), Hamiltonian graph theory~\cite{DBLP:journals/dm/HoedeV81}, and geometric model simplification~\cite{garland1997surface}.

We consider the \XFC{H} problem: Given a graph $G$ and an integer $k$, find 
whether $G$ can be transformed, by at most $k$ edge contractions, into a graph without any induced copies 
of $H$. 
The parameter we consider is $k$.
Unlike graph contraction problems, other major graph modification
problems are well-understood for these target graph classes. In particular,
P versus \NPC\ dichotomies are known for \HEE, \HED, \HEC~\cite{AravindSS17}, and \HVD~\cite{DBLP:journals/jcss/LewisY80}  (here, the allowed operations are edge editing, edge deletion, edge completion, and vertex deletion respectively). 
It is also known that all these problems are in FPT for every graph $H$~\cite{DBLP:journals/ipl/Cai96}.
The picture is far from complete for \XFC{H}.
See Table~\ref{table:pnp}.
It is trivial to note that \XFC{H} is polynomial-time solvable if $H$
is a complete graph of at most 2 vertices. 
Cai, Guo~\cite{DBLP:conf/iwpec/CaiG13, DBLP:phd/ndltd/Guo13}, and Lokshtanov, Misra, and Saurabh~\cite{DBLP:conf/iwpec/LokshtanovMS13} proved the following results for \XFC{H}.
\begin{itemize}
    \item FPT when $H$ is a complete graph
    \item If $H$ is a path or a cycle, then the problem is FPT when $H$ has at most 3 edges, and W[2]-hard otherwise.
    \item W[2]-hard when $H$ is 3-connected but not complete, or a star graph on at least 5 vertices, or a diamond.
\end{itemize}
The W[2]-hardness results mentioned above also imply NP-completeness of the problems. Guo~\cite{DBLP:phd/ndltd/Guo13} proved that
the problem is \NPC\ when $H$ is a complete graph on $t$ vertices, for every $t\geq 3$. 
Eppstein~\cite{DBLP:journals/jgaa/Eppstein09} proved that the Hadwiger number problem (find whether the size of a largest clique minor of a graph is at least $k$) is \NPC. This problem is essentially \XFC{2K_1}, if we ignore the parameter. 
This result implies that the problem is \NPC\ when $H$ is a $P_3$. We build on these results and prove the following.
\begin{itemize}
    \item \XFC{H} is \NPC\ if $H$ is not a complete graph on at most 2 vertices.
    \item \XFC{H} is W[2]-hard if $H$ is a tree which is neither a star on at most 4 vertices (\begin{tikzpicture}[myv/.style={circle, draw, inner sep=1.5pt}]
    \node[myv] (w) at (0,0) {};
\end{tikzpicture}
,\begin{tikzpicture}[myv/.style={circle, draw, inner sep=1.5pt}]
    \node (o) at (0,0) {};
    \node[myv] (v1) at (0,-0.15) {};
    \node[myv] (v2) at (0,0.15) {};
    \draw (v1) -- (v2);
\end{tikzpicture}
,\begin{tikzpicture}[myv/.style={circle, draw, inner sep=1.5pt}]
    \node[myv] (o) at (0,0) {};
    \node[myv] (v1) at (-0.15,-0.3) {};
    \node[myv] (v2) at (0.15,-0.3) {};
    \draw (o) -- (v1);
    \draw (o) -- (v2);
\end{tikzpicture}
,\begin{tikzpicture}[myv/.style={circle, draw, inner sep=1.5pt}]
    \node[myv] (o) at (0,0) {};
    \node[myv] (v1) at (-0.3,-0.3) {};
    \node[myv] (v2) at (0,-0.3) {};
    \node[myv] (v3) at (0.3,-0.3) {};
    \draw (o) -- (v1);
    \draw (o) -- (v2);
    \draw (o) -- (v3);
\end{tikzpicture}
) nor a bistar in $\{$\begin{tikzpicture}[myv/.style={circle, draw, inner sep=1.5pt}]
    \node (o) at (0,0) {};
    \node[myv] (v) at (-0.15,0) {};
    \node[myv] (vd) at (0.15,0) {};
    \node[myv] (v1) at (-0.15,-0.3) {};
    \node[myv] (vd1) at (0.15,-0.3) {};

    \draw (v) -- (vd);
    \draw (v) -- (v1);
    \draw (vd) -- (vd1);
\end{tikzpicture}
,\begin{tikzpicture}[myv/.style={circle, draw, inner sep=1.5pt}]
    \node (o) at (0,0) {};
    \node[myv] (v) at (-0.15,0) {};
    \node[myv] (vd) at (0.15,0) {};
    \node[myv] (v1) at (0,-0.3) {};
    \node[myv] (v2) at (-0.3,-0.3) {};
    \node[myv] (vd1) at (0.3,-0.3) {};

    \draw (v) -- (vd);
    \draw (v) -- (v1);
    \draw (v) -- (v2);
    \draw (vd) -- (vd1);
\end{tikzpicture}
,\begin{tikzpicture}[myv/.style={circle, draw, inner sep=1.5pt}]
    \node (o) at (0,0) {};
    \node[myv] (v) at (-0.15,0) {};
    \node[myv] (vd) at (0.15,0) {};
    \node[myv] (v1) at (-0.15,-0.3) {};
    \node[myv] (v2) at (-0.45,-0.3) {};
    \node[myv] (vd1) at (0.15,-0.3) {};
    \node[myv] (vd2) at (0.45,-0.3) {};

    \draw (v) -- (vd);
    \draw (v) -- (v1);
    \draw (v) -- (v2);
    \draw (vd) -- (vd1);
    \draw (vd) -- (vd2);
\end{tikzpicture}
$\}$. 
\end{itemize}
Our W[2]-hardness results, along with known positive results, leaves behind only three open cases among trees - \begin{tikzpicture}[myv/.style={circle, draw, inner sep=1.5pt}]
    \node[myv] (o) at (0,0) {};
    \node[myv] (v1) at (-0.3,-0.3) {};
    \node[myv] (v2) at (0,-0.3) {};
    \node[myv] (v3) at (0.3,-0.3) {};
    \draw (o) -- (v1);
    \draw (o) -- (v2);
    \draw (o) -- (v3);
\end{tikzpicture}
, \begin{tikzpicture}[myv/.style={circle, draw, inner sep=1.5pt}]
    \node (o) at (0,0) {};
    \node[myv] (v) at (-0.15,0) {};
    \node[myv] (vd) at (0.15,0) {};
    \node[myv] (v1) at (0,-0.3) {};
    \node[myv] (v2) at (-0.3,-0.3) {};
    \node[myv] (vd1) at (0.3,-0.3) {};

    \draw (v) -- (vd);
    \draw (v) -- (v1);
    \draw (v) -- (v2);
    \draw (vd) -- (vd1);
\end{tikzpicture}
, \begin{tikzpicture}[myv/.style={circle, draw, inner sep=1.5pt}]
    \node (o) at (0,0) {};
    \node[myv] (v) at (-0.15,0) {};
    \node[myv] (vd) at (0.15,0) {};
    \node[myv] (v1) at (-0.15,-0.3) {};
    \node[myv] (v2) at (-0.45,-0.3) {};
    \node[myv] (vd1) at (0.15,-0.3) {};
    \node[myv] (vd2) at (0.45,-0.3) {};

    \draw (v) -- (vd);
    \draw (v) -- (v1);
    \draw (v) -- (v2);
    \draw (vd) -- (vd1);
    \draw (vd) -- (vd2);
\end{tikzpicture}
.
\begin{table} 
\centering
\begin{tabular}{|m{2.5cm}|m{2.1cm}|m{1.9cm}|m{2.5cm}|m{3.0cm}|}
\hline
\multicolumn{1}{|c|}{\textbf{Problem}}  & \multicolumn{1}{|c|}{\textbf{P}} & \multicolumn{1}{|c|}{\textbf{NPC}} & \multicolumn{1}{|c|}{\textbf{FPT}} & \multicolumn{1}{|c|}{\textbf{W-hard}}\\ \hline
\hline
Edge Editing & $n\leq 2$~[trivial] & otherwise~\cite{AravindSS17} & For all $H$~\cite{DBLP:journals/ipl/Cai96} & \\ \hline
Edge Deletion & $m\leq 1$~[trivial] & otherwise~\cite{AravindSS17} & For all $H$~\cite{DBLP:journals/ipl/Cai96} & \\ \hline
Edge Completion & $m'\leq 1$~[trivial] & otherwise~\cite{AravindSS17} & For all $H$~\cite{DBLP:journals/ipl/Cai96} & \\ \hline
Vertex Deletion & $n\leq 1$~[trivial] & otherwise~\cite{DBLP:journals/jcss/LewisY80} & For all $H$~\cite{DBLP:journals/ipl/Cai96} & \\ \hline
$\vcenter{Edge Contraction}$ & $K_1,K_2$~[trivial] & otherwise [\textbf{Theorem \ref{thm:npc}}] & $K_t$ ($t\geq 3$)~\cite{DBLP:phd/ndltd/Guo13,DBLP:conf/iwpec/LokshtanovMS13}, $P_3, P_4, K_2+K_1$(~\cite{DBLP:conf/iwpec/CaiG13,DBLP:conf/iwpec/LokshtanovMS13}, \MSOO\ expressibility) & W[2]-hard for 3-connected non-complete graphs, diamond \cite{DBLP:conf/iwpec/CaiG13,DBLP:phd/ndltd/Guo13}, $C_t$ ($t\geq 4$)~\cite{DBLP:conf/iwpec/CaiG13,DBLP:conf/iwpec/LokshtanovMS13}, all trees except 7 trees~[\textbf{Theorem~\ref{thm:whard}}]\\ \hline
\end{tabular}
\caption{Complexities of various graph modification problems where the target property is $H$-free. The number of vertices, the number of edges, and the number of nonedges in $H$ are denoted by $n,m,m'$ respectively.}
\label{table:pnp}
\end{table}
\section{Preliminaries}
\label{sec:prelims}
\subsubsection*{Graphs.}
All graphs considered in this paper are simple and undirected.
A complete graph and a path on $t$ vertices are denoted by $K_t$
and $P_t$ respectively.
A \textit{universal vertex} of a graph is a vertex adjacent to every other
vertex of the graph. An \textit{isolated} vertex is a vertex 
with degree 0.
For an integer $t\geq 0$,
a \textit{star} on $t+1$ vertices, denoted by $K_{1,t}$, is a tree
with a single universal vertex and $t$ degree-1 vertices.
The universal vertex in $K_{1,t}$ is also called the \textit{center} of the star.
For integers $t,t'$ such that $t\geq t'\geq 0$, a \textit{bistar} on $t+t'+2$ vertices, denoted by $T_{t,t'}$, is a 
tree with two adjacent vertices $v$ and $v'$, where $t$
degree-1 vertices are attached to $v$ and $t'$ degree-1 
vertices are attached to $v'$. The bistar $T_{t,0}$ is the star $K_{1,t+1}$ and the bistar $T_{1,1}$ is $P_4$.
We say that $vv'$ is the \textit{central edge} of the bistar.
By $G_1+G_2$ we denote the disjoint union of the graphs $G_1$ and $G_2$.
A graph is $H$-free if it does not contain any induced copies of $H$.
In a graph $G$, \textit{replacing a vertex} $v$ with a graph $H$ is the 
graph obtained from $G$ by removing $v$, introducing a copy of $H$, and adding edges
between every vertex of the $H$ and every neighbor of $v$ in $G$.
A \textit{separator} $S$ of a connected graph $G$ is a subset of its vertices
such that $G-S$ (the graph obtained from $G$ by removing the vertices in $S$) is disconnected. A separator is \textit{universal} if every
vertex of the separator is adjacent to every vertex outside the separator. We will be using the term `universal $K_1$ (resp. $K_2$) separator' to denote a universal separator which induces a $K_1$ (resp. $K_2$). 
Let $V'$ be a subset of vertices of a graph $H$. By $H[V']$ we 
denote the graph induced by $V'$ in $H$.
For a graph $G$ and two subsets $A$ and $B$ of vertices of $G$,
by $E[A,B]$ we denote the set of edges in $G$, where each edge in the set is having one end 
point in $A$ and the other end point in $B$.

\subsubsection*{Contraction.}
\textit{Contracting} an edge $uv$ in a graph $G$ is the operation in which the vertices $u$ and $v$ are identified to be a new vertex $w$ such 
that $w$ is adjacent to every vertex adjacent to either $u$ or $v$.
Given a graph $G$ and a subset $F$ of edges of $G$, the graph $G/F$ 
obtained by contracting the edges in $F$ does not depend on the order in which the edges are contracted.
Every vertex $w$ in $G/F$ represents a subset $W$ of vertices (which are contracted to $w$) of $G$
such that $W$ induces a connected graph in $G$. 
Let $G_F$ be the subgraph of $G$ containing all vertices of $G$
and the edges in $F$. There is a partition $\mathcal{P}$
of vertices of $G$ implied by $F$: Every set in $\mathcal{P}$
corresponds to the vertices of a connected component in $G_F$.
We note that
many subsets of edges may imply the same partition - it does not matter which all edges of a connected subgraph are contracted to get a single vertex. 
The graph $G/F$ is nothing but the graph in which there is a vertex corresponding to every set in $\mathcal{P}$ and two vertices in $G/F$
are adjacent if and only if there is at least one edge in $G$ between the corresponding sets in $\mathcal{P}$.
The graph $G/F$ is equivalently denoted by $G/\mathcal{P}$.
Assume that $\mathcal{P}'$ is 
a partition of a subset of vertices of $G$.
Then by $G/\mathcal{P}'$ we denote the graph 
obtained from $G$ by contracting each set in $\mathcal{P}'$
into a single vertex.
The \textit{cost} of a set $P$ in $\mathcal{P}$ is the number $|P|-1$, which is equal to the minimum number of edges required to \textit{form} the set $P$. The cost of $\mathcal{P}$, denoted by \SURP, is the sum of costs of the sets in $\mathcal{P}$. We observe that $|\mathcal{P}|+$\SURP\ $= n$, where $n$ is the number of vertices of $G$.
We say that $F$ \textit{touches} a subset $W$ of vertices of $G$, if there 
is at least one edge $uv$ in $F$ such that either $u$ or $v$ is in $W$.
Let $u,v$ be two non-adjacent vertices of $G$. \textit{Identifying}
$u$ and $v$ in $G$ is the operation of removing $u$ and $v$,
adding a new vertex $w$, and making $w$ adjacent to every vertex
adjacent to either $u$ or $v$. 

\subsubsection*{Fixed parameter (in)tractability.}
A parameterized problem is fixed-parameter tractable (FPT) if it can
be solved in time $f(k)|I|^{O(1)}$-time, where $f$ is a computable
function and $(I,k)$ is the input.
Parameterized problems fall into different levels of complexities which are captured by the W-hierarchy. 
A parameterized reduction from a parameterized problem $Q'$
to a parameterized problem $Q$ is an algorithm which takes
as input an instance $(I',k')$ of $Q'$ and outputs an instance 
$(I,k)$ of $Q$ such that the algorithm runs in time $f(k')|I'|^{O(1)}$ (where $f$ is a computable function), and
$(I',k')$ is a yes-instance of $Q'$ if and only if $(I,k)$
is a yes-instance of $Q$,
and $k\leq g(k')$ (for a computable function $g$).
We use parameterized reductions to 
transfer fixed-parameter intractability. For more details on these topics, we 
refer to the textbook \cite{DBLP:books/sp/CyganFKLMPPS15}.
The problem that we deal with in this paper is defined as follows.
\begin{mdframed}
  \XFC{H}:
  Given a graph $G$ and an integer $k$, can $G$ be modified into an $H$-free graph by at most $k$ edge contractions?
\end{mdframed}

\section{NP-completeness}
In this section we prove that \XFC{H} is \NPC\ whenever $H$ is not a complete graph of at most two vertices. 
In Section~\ref{sub:npc-general} we obtain reductions for the cases when $H$ is connected but
does not have any universal $K_1$ separator and universal $K_2$ separator. In Section~\ref{sub:npc-uni} we deal with non-star graphs with universal $K_1$ separator or universal $K_2$ separator. In Section~\ref{sub:npc-stars} we resolve the stars. The cases when
$H$ is a $2K_2$ or a $K_2+K_1$ are handled in Section~\ref{sub:npc-small}. These come as base cases in the inductive proof of the main result of the section given in Section~\ref{sub:npc-ptt}.
We crucially use the following results.
\begin{proposition}[{\cite{DBLP:phd/ndltd/Guo13,DBLP:journals/jgaa/Eppstein09}}]
    \label{pro:npc-known}
    \XFC{H} is \NPC\ when $H$ is a $2K_1$, or a $P_3$, or a $K_t$,
    for any $t\geq 3$.
\end{proposition}

\label{sec:npc}
\subsection{A general reduction}
\label{sub:npc-general}
\begin{figure}
    \centering
    \begin{tikzpicture}[myv/.style={circle, draw, inner sep=1.5pt}]
    \node[myv,label={[blue]87:$w$}] (w) at (0,0) {};
    \node[myv,fill=black,label={[blue]90:$z$}] (z) at (90:1.2) {};
    \node[myv,fill=black,label={[blue]210:$x$}] (x) at (210:1.2) {};
    \node[myv,fill=black,label={[blue]330:$y$}] (y) at (330:1.2) {};
    \node[myv] (xz2) at (150:1.2) {};
    \node[myv] (xz1) at (150:1.5) {};
    \node[myv] (xz3) at (150:0.9) {};
    \node[myv] (xy2) at (270:1.2) {};
    \node[myv] (xy1) at (270:1.5) {};
    \node[myv] (xy3) at (270:0.9) {};
    \node[myv] (yz2) at (30:1.2) {};
    \node[myv] (yz1) at (30:1.5) {};
    \node[myv] (yz3) at (30:0.9) {};
    \draw (w) -- (xy3);    
    \draw (w) -- (xz3);    
    \draw (w) -- (yz3);    
    \draw (xz1) -- (xz2);    
    \draw (xz2) -- (xz3);    
    \draw (xy1) -- (xy2);    
    \draw (xy2) -- (xy3);    
    \draw (yz1) -- (yz2);    
    \draw (yz2) -- (yz3);    
    \draw (x) -- (xy1);    
    \draw (x) -- (xy2);    
    \draw (x) -- (xy3);    
    \draw (x) -- (xz1);    
    \draw (x) -- (xz2);    
    \draw (x) -- (xz3);    
    \draw (y) -- (xy1);    
    \draw (y) -- (xy2);    
    \draw (y) -- (xy3);    
    \draw (y) -- (yz1);    
    \draw (y) -- (yz2);    
    \draw (y) -- (yz3);    
    \draw (z) -- (yz1);    
    \draw (z) -- (yz2);    
    \draw (z) -- (yz3);    
    \draw (z) -- (xz1);    
    \draw (z) -- (xz2);    
    \draw (z) -- (xz3);    
    \draw (w) -- (x);    
    \draw (w) -- (y);    
    \draw (w) -- (z);    

    \draw [dashed, red, rotate around={150:(xz2)}] (150:1.2) ellipse (0.75 and 0.5);
    \draw [dashed, red, rotate around={270:(xy2)}] (270:1.2) ellipse (0.75 and 0.5);
    \draw [dashed, red, rotate around={30:(yz2)}] (30:1.2) ellipse (0.75 and 0.5);
    \node[label={[blue]150:$W_{xz}$}] at (150:1.7) {};
    \node[label={[blue]30:$W_{yz}$}] at (30:1.6) {};
    \node[label={[blue]270:$W_{xy}$}] at (270:1.8) {};

    \end{tikzpicture}
    \caption{Construction of $G$ from $(G',H)$ by Construction~\ref{con:vc}, where $G'$ is a triangle and $H$ is a $P_4$. The vertices of $G'$ in $G$ are darkened.}
    \label{fig:vc}
\end{figure}
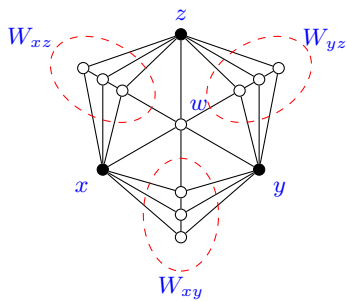

A vertex cover of a graph $G$ is a subset $V'$ of its vertices such that for every edge $uv$ of $G$, either $u$ or $v$ is in $V'$.
\VC\ is the decision problem in which the objective is to check
whether the given graph has a vertex cover of size at most $k$.
The problem is very well-known to be an \NPC\ problem.
Inspired by a reduction by Asano and Hirata~\cite{asano1983edge}, we introduce the following reduction from \VC\  which handles connected graphs $H$ without any universal $K_1$ separator and universal $K_2$ separator.
\begin{construction}
\label{con:vc}
Let $G'$ be a graph without any isolated vertices, and $H$ be a connected non-complete graph. We obtain a graph $G$ from $G'$ and $H$ as follows.
\begin{itemize}
    \item Subdivide each edge of $G'$ once, i.e., for every edge $uv$,
    introduce a new vertex and make it adjacent to both
    $u$ and $v$, and delete the edge $uv$.
    \item Replace each new vertex by a copy $H$. 
    \item Let $w$ be a non-universal vertex of $H$ (the existence of $w$ is guaranteed as $H$ is not a complete graph). Identify $w$ of every copy of $H$ (introduced in the previous step) to be a single vertex named $w$.   
\end{itemize}
Let the resultant graph be $G$. The vertices in $G$ copied from $G'$ form the set $V'$, which forms an independent set in $G$. For each edge $uv$ in $G'$, the vertices, except $w$, of the copy of $H$ is denoted by $W_{uv}$. By $W$ we denote any such set. We note that $w$ is adjacent to every vertex in $V'$, as $G'$ does not have any isolated vertices. This completes the construction. An example is shown in Figure~\ref{fig:vc}.
\end{construction}

Let $H$ be a connected non-complete graph with $h$ vertices and without any universal $K_1$ or $K_2$ separator. Let $G'$ be a graph without any isolated vertices. Let $G$ be obtained from $(G',H)$ by Construction~\ref{con:vc}. 
Assume that $(G',k)$ is a yes-instance of \VC\ and let $T$ be any vertex cover of size at most $k$ of $G'$. Let $F = \{wu: u\in T\}$.
Let $w$ itself denote the vertex obtained by contracting the edges in $F$. 
The following can be observed directly from the fact that $T$ is a 
vertex cover.
\begin{observation}
\label{obs:vc}
The vertex $w$ is a universal vertex in $G/F$.
\end{observation}
Before, proving that $G/F$ is $H$-free, we prove the following lemma.

\begin{lemma}
    \label{lem:vc:forward-case}
    Let $v$ be a vertex in $V'\setminus T$. Let $u$ be any neighbor of $v$ in $G'$. Then none of the subsets of $W_{uv}\cup\{v,w\}$ induces an $H$ in $G/F$. 
\end{lemma}
\begin{proof}
    For a contradiction, assume that a subset $U$ of $W_{uv}\cup \{v,w\}$ induces an $H$ in $G/F$. Since $w$ is not a universal 
    vertex in $H$, and both $w$ (by Observation~\ref{obs:vc}) and $v$ (by the construction) are universal to $W_{uv}$
    in $G/F$, we obtain that neither $W_{uv}\cup \{v\}$ nor $W_{uv}\cup \{w\}$ induces $H$ in $G/F$. Therefore, $U=(W_{uv}\setminus \{x\})
    \cup \{v,w\}$, for some vertex $x$ in $W_{uv}$. Let $m$ be the number of edges in $H$. Let $m'$ be the number of edges in the graph
    induced by $W_{uv}\setminus \{x\}$ in $H$. Clearly, $m = m'+d_x+d_w-d_{wx}$, where $d_x$ and $d_w$ denote the degree of $x$ and degree of $w$ respectively in $H$, and $d_{wx}$ is 1 if there is an edge between $w$ and $x$ in $H$, and 0 otherwise. Since $U$ induces 
    an $H$ in $G/F$, we obtain that $m = m'+h-1+h-1-1 = m'+2h-3$ (note
    that both $w$ and $v$ are universal in the $H$ induced by $U$).
    Therefore, we obtain that $d_x+d_w-d_{xw} = 2h-3$. Since $w$
    is not universal in $H$, we obtain that $d_w\leq h-2$. Therefore,
    $d_x-d_{xw}\geq h-1$. If $x$ is universal in $H$, then we get that
    $d_x-d_{xw}= h-1-1=h-2$, and if $x$ is not universal in $H$, then we obtain that
    $d_x-d_{xw}\leq h-2-0 = h-2$. Therefore, we get a contradiction.    
\end{proof}
\begin{lemma}
\label{lem:vc-forward}
If $(G',k)$ is a yes-instance of \VC, then $(G,k)$ is a yes-instance of \XFC{H}.
\end{lemma}
\begin{proof}
Let $T$ be a vertex cover of size at most $k$ of $G'$ 
and $F = \{wu: u\in T\}$. We claim that $G/F$ is $H$-free. For a 
contradiction, assume that there is a copy of $H$ induced by a set $U$ 
in $G/F$. 
By Observation~\ref{obs:vc}, $w$ is 
a universal vertex in $G/F$. Since the vertices in $V'\setminus T$
along with $w$ induces a star graph
in $G/F$, and $H$ is a not a
subgraph of a star graph
(a non-complete star graph has a universal $K_1$ separator), 
$U$ cannot be 
a subset of $(V'\setminus T)\cup\{w\}$. Therefore, $U$
must include some vertices from at least one copy of $W$, say $W_{uv}$. Since $w$
is universal to $W_{uv}$ and $w$ is not a universal vertex in $H$,
we obtain that $W_{uv}\cup \{w\}$ does not induce an $H$.
Since $T$ is a vertex cover, either $u$ or $v$ is in $T$. 
Assume that both are in $T$. Then $U$ must contain at least 
one vertex not in $W_{uv}\cup \{w\}$ and then $w$ is a universal
$K_1$ separator in the $H$ induced by $U$. 
Assume that $u\in T$ and $v\notin T$. By Lemma~\ref{lem:vc:forward-case}, none of the subsets of $W_{uv}\cup\{v,w\}$ induces an $H$.
Let $W_v$ denote the union of all $W_{u'v}$ (for every edge $u'v$ in $G'$).
If $U$ contains at least one vertex not in $\{v,w\}\cup W_v$, 
then $w$ is a universal $K_1$ separator in the $H$ induced by $U$.
If $U$ does not contain any such vertex, then $U$ must contain
a vertex from $W_{u'v}$ for some $u'\neq u$. 
Then either $w$ or $v$ is a universal $K_1$ separator
or $wv$ is a universal $K_2$ separator in the $H$
induced by $U$.
Thus, in all these cases,
we obtain contradictions.
%
\end{proof}

\begin{lemma}
    \label{lem:vc:backward}
    Let $(G,k)$ be a yes-instance of \XFC{H}. Then $(G',k)$
    is a yes-instance of \VC. 
\end{lemma}
\begin{proof}
    Let $F$ be a set of edges of $G$ such that $|F|\leq k$ and
    $G/F$ is $H$-free. We create a vertex cover $T$ of $G'$ as 
    follows. 
    \begin{itemize}
        \item For every vertex $u$ of $G'$, if $wu$ is in $F$, then include $u$ in $T$.
        \item For every edge $uv$ of $G'$, if there is an edge $uv'\in F$, where $v'\in W_{uv}$, then include $u$ in $T$.
        \item For every edge $uv$ of $G'$, if there is an edge $u'v'\in F$, or an edge $wu'\in F$, where $u',v'\in W_{uv}$, include either $u$ or $v$ in $T$
        arbitrarily.
    \end{itemize}
    Clearly, $|T|\leq |F|\leq k$. Now we prove that $T$ is a vertex 
    cover of $G'$. We note that $W_{uv}\cup \{w\}$ induces an $H$ in $G$
    for every edge $uv$ in $G'$.
    Therefore, a vertex in $W_{uv}\cup \{u,v\}$
    must be touched by $F$ (the only neighbors of $W_{uv}$ are $u,v$, and $w$). Therefore, by the construction of $T$,
    either $u$ or $v$ is in $T$. Hence $T$ is a vertex cover of $G'$.
\end{proof}

NP-completeness of \VC\ and Lemmas~\ref{lem:vc-forward}, \ref{lem:vc:backward} prove Lemma~\ref{lem:npc:general}.
\begin{lemma}
\label{lem:npc:general}
Let $H$ be a connected non-complete graph with neither 
a universal $K_1$ separator nor a universal $K_2$ separator.
Then \XFC{H} is \NPC.
\end{lemma}
\subsection{Graphs with universal clique separators}
\label{sub:npc-uni}
In this section, we handle the graphs $H$ with either a universal
$K_1$ separator (except stars) or with a universal $K_2$ separator. We note that
$H$ cannot have both a universal $K_1$ separator and a universal $K_2$
separator. Further $H$ cannot have more than one such separator.

\begin{construction}
\label{con:copy1}
Let $G', H$ be any graphs and let $V'$ be any subset of vertices of $H$. Let $b,c,k$ be positive integers. 
We obtain a graph $G$ from $(G',H,V',b,c,k)$ as follows. For every set $S$ of vertices of $G'$, where $S$ induces a clique on $b$ vertices in $G'$, do the following: Introduce $k+c$ copies of $H[V']$ and make every vertex of the copies adjacent to every vertex of $S$. 
Let $W_S$ denote the set of new vertices introduced for $S$, and 
let $W$ be the set of all new vertices.
\end{construction}
\begin{figure}
    \centering
    \begin{tikzpicture}[myv/.style={circle, draw, inner sep=1.5pt}]
    \node (w) at (0,0) {};
    \node[myv,fill=black] (z) at (90:1) {};
    \node[myv,fill=black] (x) at (210:1) {};
    \node[myv,fill=black] (y) at (330:1) {};
    \node[myv] (x1) at (205:1.5) {};
    \node[myv] (x2) at (215:1.5) {};
    \node[myv] (y1) at (325:1.5) {};
    \node[myv] (y2) at (335:1.5) {};
    \node[myv] (z1) at (85:1.5) {};
    \node[myv] (z2) at (95:1.5) {};
    \draw (x) -- (y);
    \draw (x) -- (z);
    \draw (z) -- (y);
    \draw (x) -- (x1);
    \draw (x) -- (x2);
    \draw (y) -- (y1);
    \draw (y) -- (y2);
    \draw (z) -- (z1);
    \draw (z) -- (z2);
    \end{tikzpicture}
    \caption{Construction of $G$ from $(G',H, V', b,c,k)$ by Construction~\ref{con:copy1}, 
    where $G'$ is a triangle and $H$ is a paw. 
    Since paw has a universal $K_1$ separator 
    (denote it by $K$), $b=|K|=1$. Since we get a $K_1$ and a 
    $K_2$ after removing 
    the universal $K_1$ separator from paw, the smallest 
    component is $K_1$. Therefore $V'$ contains a single vertex. 
    Since there are only one copy of $K_1$ (as a component) in $H-K$, $c=1$. 
    Assume that $k=1$. The vertices of $G'$ in $G$ are darkened.}
    \label{fig:unisep1}
\end{figure}
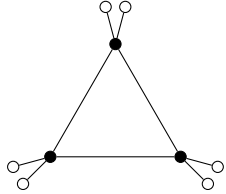

Let $H$ be a graph with a universal $K_1$ separator or a universal $K_2$ separator. Let the set of vertices of the separator be denoted by $K$. Assume that there exists at least two non-isomorphic components
in $H-K$ (therefore, $H$ cannot be a star). Let $J$ be a component in $H-K$ with minimum number of 
vertices. Let $c$ be the number of times $J$ appears (as a component) in $H-K$.
Let $V'$ be the set of vertices of a copy of $J$.
Let $H'$ be 
the graph obtained from $H$ by removing the vertices of every component isomorphic to $J$ in $H-K$. 
Let $(G',k)$ be an instance of \XFC{H'}.
Let $G$ be obtained
from $(G',H,V',b=|K|,c,k)$ by Construction~\ref{con:copy1}.  An example of the construction is shown in Figure~\ref{fig:unisep1}.

\begin{lemma}
    \label{lem:unisep1:forward}
    Let $(G',k)$ be a yes-instance of \XFC{H'}. Then $(G,k)$
    is a yes-instance of \XFC{H}.
\end{lemma}
\begin{proof}
Let $\mathcal{P}'$ be a partition of vertices of $G'$ such that
each set induces a connected graph, \SURPD$\leq k$, and $G/\mathcal{P}'$ is $H'$-free.
Let $\mathcal{P}$ be obtained from $\mathcal{P}'$ by
adding the singleton sets corresponding to vertices in $W$.
Let $\mathcal{P}_W$ be the singleton sets corresponding to vertices in $W$, i.e., $\mathcal{P} = \mathcal{P}'\cup \mathcal{P}_W$.
Clearly, \SURP$=$\SURPD$\leq k$.
We claim that $G/\mathcal{P}$ is $H$-free.
For a contradiction, assume that $\mathcal{U}\subseteq \mathcal{P}$
induces an $H$ in $G/\mathcal{P}$.
Since there are at least two components in $H-K$, we obtain that
degree of a vertex in $J$ is strictly less than that of a vertex in
$K$ in $H$. Therefore, None of the sets in $\mathcal{P}_W$
can act as a universal vertex (a vertex in $K$) in the $H$ induced by
$\mathcal{U}$. 
Assume that $|K|=1$. Let $P_u\in \mathcal{P'}$ be the universal
vertex of the induced $H$. Since $J$ is a component in $H-K$
with minimum number of vertices, if a singleton set 
corresponding to a vertex in $W_{\{u\}}$ 
(for any vertex $u\in P_u$) is in $\mathcal{U}$, then each of the singleton
sets corresponding to vertices in $W_{\{u\}}$ (forming a copy of $J$) is in $\mathcal{U}$.
Among the sets in $P_W$, the set $P_u$ is adjacent to only the singleton
sets corresponding to vertices in $W_{\{u\}}$ (for $u\in P_u$). 
Therefore, the singleton sets corresponding to $W_v$, for $v\notin P_u$, are not in $\mathcal{U}$. Therefore, there must be an 
    induced copy of $H'$ in $G'/\mathcal{P}'$, which is a contradiction.
    The case when $|K|=2$ can be proved in a similar way.
\end{proof}

\begin{lemma}
\label{lem:unisep1:backward}
Let $(G,k)$ be a yes-instance of \XFC{H}. Then $(G',k)$ is a yes-instance of \XFC{H'}.
\end{lemma}
\begin{proof}
Let $F$ be a subset of edges of $G$ such that $G/F$ is $H$-free 
and $|F|\leq k$. 
Let $\mathcal{P}$ be the partition of vertices 
of $G$ corresponding to $F$.
We create a partition $\mathcal{P}'$ from $\mathcal{P}$ as follows. 
For every $P\in \mathcal{P}$, add the set $P\setminus W$ to $\mathcal{P}'$. Clearly, $\mathcal{P}'$ is a partition of vertices of $G'$ and each set in $\mathcal{P}'$ induces a connected subgraph in
$G'$. Further, \SURPD\ $\leq$ \SURP\ $\leq k$. Now, it is sufficient
to prove that $G'/\mathcal{P}'$ is $H'$-free.
For a contradiction, assume that $G'/\mathcal{P}'$ has an $H'$
induced by $\mathcal{U}\subseteq \mathcal{P}'$. 

Let $|K|=1$. Let $P'\in \mathcal{U}$ be
a set corresponding to the universal $K_1$ separator (of $H$) which is part of the induced $H'$. Consider any vertex $u\in P'$. Since there are $k+c$
copies of $J$ attached to $u$, at least $c$ of them are not touched by $\mathcal{P}$ (i.e., each vertex of those $c$ copies of $J$ forms a singleton set in $\mathcal{P}$). Therefore, the singleton sets in $\mathcal{P}$ corresponding the $c$ copies of $J$ along with  the sets in $\mathcal{P}$ containing sets in $\mathcal{U}$ induce an $H$ in $G/\mathcal{P}$, which is a contradiction. 

With similar arguments we can handle the case of $|K|=2$. Let $|K|=2$.
Let $P_u', P_v'\in \mathcal{U}$ be
two sets corresponding to the two vertices in the universal $K_2$ separator (of $H$), which is part of the induced $H'$. 
Since $P_u'$ and $P_v'$ are adjacent, there exists a vertex $u\in P_u'$
and a vertex $v\in P_v'$ such that $uv$ is an edge in $G'$. Since there are $k+c$
copies of $J$ attached to $uv$, at least $c$ of them are not touched by $\mathcal{P}$ (i.e., each vertex of those $c$ copies of $J$ forms a singleton set in $\mathcal{P}$). Therefore, the singleton sets in $\mathcal{P}$ corresponding the $c$ copies of $J$ along with the sets in $\mathcal{P}$ containing sets in $\mathcal{U}$ induce an $H$ in $G/\mathcal{P}$, which is a contradiction.
\end{proof}

We obtain Lemma~\ref{lem:unisep1} from Lemmas~\ref{lem:unisep1:forward} and \ref{lem:unisep1:backward}.
\begin{lemma}
    \label{lem:unisep1}
    Let $H$ be a graph with a universal $K_1$ separator
    or a universal $K_2$ separator, denoted by $K$. 
    Assume that $H-K$ has at least two components which are
    not isomorphic. Let $J$ be a component in $H-K$ with minimum
    number of vertices. Let $H'$ be obtained from $H$
    by removing all components of $H-K$ isomorphic to $J$.
    Then there is a polynomial-time reduction from \XFC{H'}
    to \XFC{H}.
\end{lemma}

What remains to handle is the case when $H$ has a universal $K_1$
separator or a universal $K_2$ separator $K$ such that $H-K$
is a disjoint union of a graph $J$. The diamond graph 
is an example.
For this we need the concept of an enforcer - a structure to forbid
contraction of certain edges. Enforcers are used widely in connection with proving hardness results for edge modification problems (see \cite{CaiC15incompressibility,DBLP:phd/ndltd/Guo13,DBLP:journals/jcss/MarxS22}). 

We define an \textit{enforcer} of a graph $H$ as a graph $H_E$ with two specified adjacent vertices $v,w$ such that the following hold true for
every instance $(G,k)$ of \XFC{H} and for every edge $ab$ of $G$:
$(\hat{G},k)$ is a yes-instance of \XFC{H} if and only if
there is a partition $\mathcal{P}$ of vertices of $G$ (where
each set in $\mathcal{P}$ induces a connected subgraph of $G$) such that $G/\mathcal{P}$ is 
$H$-free, \SURP$\leq k$, and $a$ and $b$ are in different sets in $\mathcal{P}$. Here, $\hat{G}$ is the graph 
obtained from $G$ by attaching $k+1$ copies of $H_E$ by 
identifying $a$ with $v$, and $b$ with $w$ for each copy of $H_E$.

\begin{figure}
  \centering
  \begin{subfigure}[b]{0.2\textwidth}
    \centering
    \begin{tikzpicture}[myv/.style={circle, draw, inner sep=1.5pt}]
    \node[myv,label={[blue]0:$u$}] (u) at (0,0) {};
    \node[myv,label={[blue]0:$v$}] (v) at (60:1) {};
    \node[myv] (uv) at (120:1) {};
    \node[myv] (vd) at (300:1) {};
    \node[myv] (uvd) at (240:1) {};

    \draw (u) -- (v);
    \draw (u) -- (uv);
    \draw (u) -- (vd);
    \draw (u) -- (uvd);
    \draw (v) -- (uv);
    \draw (vd) -- (uvd);
\end{tikzpicture}
    \caption{$H$}
    \label{fig:enfH}
  \end{subfigure}%
  \begin{subfigure}[b]{0.2\textwidth}
    \centering
    \begin{tikzpicture}[myv/.style={circle, draw, inner sep=1.5pt}]
    \node[myv] (u) at (0,0) {};
    \node[myv,label={[blue]0:$v$}] (v) at (60:1) {};
    \node[myv,label={[blue]0:$w$}] (w) at (60:0.5) {};
    \node[myv] (uv) at (120:1) {};
    \node[myv] (vd) at (300:1) {};
    \node[myv] (uvd) at (240:1) {};

    \draw (u) -- (w);
    \draw (v) -- (w);
    \draw (u) -- (uv);
    \draw (u) -- (vd);
    \draw (u) -- (uvd);
    \draw (v) -- (uv);
    \draw (vd) -- (uvd);
\end{tikzpicture}
    \caption{$H_E$}
    \label{fig:enfHE}
  \end{subfigure}%
  \begin{subfigure}[b]{0.5\textwidth}
    \centering
    \begin{tikzpicture}[myv/.style={circle, draw, inner sep=1.5pt}]
    \node[myv,label={[blue]-45:$a$},fill=black] (a) at (0,0) {};
    \node[myv,label={[blue]-45:$b$},fill=black] (b) at (0,-0.8) {};
    \node[myv,fill=black] (c) at (0,-1.6) {};
    \node[myv,fill=black] (d) at (0,0.8) {};
    \node[myv] (e) at (-0.8,0) {};
    \node[myv] (f) at (-0.8,-0.8) {};
    \node[myv] (g) at (-1.6,-0.8) {};
    \node[myv] (h) at (-0.8,-1.6) {};
    \node[myv] (i) at (0.8,0) {};
    \node[myv] (j) at (0.8,-0.8) {};
    \node[myv] (k) at (1.6,-0.8) {};
    \node[myv] (l) at (0.8,-1.6) {};

    \draw (c) -- (b);
    \draw (a) -- (b);
    \draw (a) -- (d);
    \draw (a) -- (e);
    \draw (a) -- (i);
    \draw (f) -- (b);
    \draw (b) -- (j);
    \draw (e) -- (f);
    \draw (i) -- (j);
    \draw (j) -- (l);
    \draw (l) -- (k);
    \draw (j) -- (k);
    \draw (f) -- (g);
    \draw (g) -- (h);
    \draw (f) -- (h);
\end{tikzpicture}
    \caption{$\hat{G}$ - the vertices of $G$ are darkened}
    \label{fig:enfgcap}
  \end{subfigure}%
  \caption{Construction of $\hat{G}$ from $(G=P_4,a,b,H_E,v,w,k=1)$}
  \label{fig:enf-all}
\end{figure}
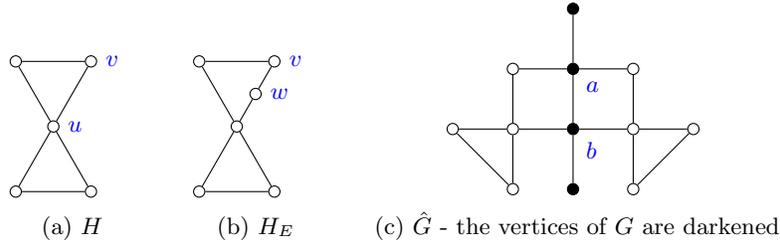

Now, we will handle the case in which $H-K$ is the disjoint union of $t\geq 2$ copies of a connected graph $J$. Assume that $H$ is not a star graph. Let $u$ be a vertex in $K$ and $v$ be a vertex in $H-K$.
Clearly, $uv$ is an edge. Subdivide $uv$ by introducing a new vertex $w$, making it adjacent to both $u$ and $v$, and deleting the edge $uv$. We note that contracting the edge $vw$ in $H_E$ creates $H$. We will prove that $H_E$ along with $v,w$ is an enforcer for $H$.

Let $\hat{G}$ be obtained from $(G,a,b,H_E,v,w, k)$ as described above.
An example is shown in Figure~\ref{fig:enf-all}.
Let $S_1, S_2,\ldots,S_{k+1}$ denote the set of vertices of 
$k+1$ copies of $H_E$, except $\{v,w\}$.
Let $S$ be the set of all new vertices, i.e., $S=\bigcup_{i=1}^{i=k+1}S_i$.
Let the number of vertices in $H$ be $h$.

\begin{lemma}
    \label{lem:unisep:enforcer:forbid}
    Let $\mathcal{\hat{P}}$ be a partition of vertices of $\hat{G}$ (where each set in $\mathcal{\hat{P}}$ induces a connected subgraph in $\hat{G}$)
    such that \SURPC$\leq k$, and $\hat{G}/\mathcal{\hat{P}}$ is $H$-free. 
    Let $P_a$ and $P_b$ denote the
    sets in $\mathcal{\hat{P}}$ containing $a$ and $b$ respectively.
    Then $P_a\neq P_b$.
\end{lemma}
\begin{proof}
    For a contradiction, assume that $P_a=P_b$. 
    There are 
    $k+1$ copies of $H$ in $\hat{G}/\{P_a\}$ where all the copies are attached
    to $P_a=P_b$.
    This implies that all of them cannot be destroyed by $\mathcal{\hat{P}}$. 
    Then there exists a set $S_i$ such that every vertex in $S_i$
    is a singleton set in $\mathcal{\hat{P}}$. Then those sets and $P_a$
    induce an $H$ in $\hat{G}/\mathcal{\hat{P}}$, which is a contradiction.
\end{proof}

\begin{lemma}
    \label{lem:unisep:enforcer:solve}
    Let $\mathcal{P}$ be a  partition of vertices of $G$ (where each set in $\mathcal{P}$ induces a connected subgraph in $G$) such that $a$ and $b$ are not in the same set in $\mathcal{P}$, and $G/\mathcal{P}$ is $H$-free. 
    Let $\mathcal{\hat{P}}$ be obtained from $\mathcal{P}$ by adding
    singleton sets corresponding to vertices in $S$.
    Then $\hat{G}/\mathcal{\hat{P}}$ is $H$-free.
\end{lemma}
\begin{proof}
For a contradiction, assume that $\mathcal{U}\subseteq \mathcal{\hat{P}}$
induces an $H$ in $\hat{G}/\mathcal{\hat{P}}$.
Clearly, none of the subsets of $\mathcal{P}$ induces an $H$ in $\hat{G}$.
Hence at least one set in $\mathcal{\hat{P}}\setminus \mathcal{P}$
is part of $\mathcal{U}$, let it be from $S_i$. 
Let $\mathcal{P}_i$ be the subset of singleton sets in $\mathcal{\hat{P}}$ corresponding to 
$S_i$. Also assume that $P_a$ and $P_b$ 
be the sets in $\mathcal{\hat{P}}$ containing $a$ and $b$ respectively. 
Clearly, $\mathcal{P}_i\cup \{P_a, P_b\}$ induces an $H_E$ and not an $H$. Therefore, $\mathcal{U}$ contains sets corresponding to $S_j$ ($j\neq i$)
or from $\mathcal{P}\setminus \{P_a,P_b\}$. 
Therefore, either $P_a$ or $P_b$ acts as a universal vertex in the $H$
induced by $\mathcal{U}$. 
Recall that $b$ is identified with $w$, a degree two vertex in $H_E$,
whose neighbors are not adjacent to each other (one of them is $v$ which is identified with $a$). Therefore, 
if $P_b$ acts as a universal vertex, then $J$ can have only one vertex.
This is a contradiction as $H$ is not a star graph.
Recall that $a$ is identified with a vertex $v$ of $H_E$ which is adjacent to at most $|V(J)|$ vertices including $w$, where $w$
is not adjacent to any neighbors of $v$. Therefore, if 
$v$ acts as a universal vertex, $J$ can have only one vertex,
which is a contradiction as $H$ is not a star graph.
\end{proof}

Lemmas~\ref{lem:unisep:enforcer:forbid} and \ref{lem:unisep:enforcer:solve} imply that $(H_E,v,w)$
is an enforcer for $H$.

\begin{lemma}
    \label{lem:enforcer}
    $(H_E,v,w)$ is an enforcer for $H$.
\end{lemma}
\begin{proof}
We need to prove that $(\hat{G},k)$ is a yes-instance of \XFC{H} if and only if
there is a partition $\mathcal{P}$ of vertices of $G$ (where
each set in $\mathcal{P}$ induces a connected subgraph of $G$) such that $G/\mathcal{P}$ is 
$H$-free, \SURP$\leq k$, and $a$ and $b$ are in different sets in $\mathcal{P}$. Lemma~\ref{lem:unisep:enforcer:solve} proves the 
backward direction, i.e., if there is a partition $\mathcal{P}$ as specified, then $(\hat{G}, k)$ is a yes-instance. For the 
forward direction, assume that $(\hat{G}, k)$ is a yes-instance.
Let $\mathcal{\hat{P}}$ be a partition of vertices of $\hat{G}$
such that \SURPC$\leq k$ and $\hat{G}/\mathcal{\hat{P}}$
is $H$-free. Let $P_a$ and $P_b$ be the sets in $\mathcal{\hat{P}}$
containing $a$ and $b$ respectively. 
By Lemma~\ref{lem:unisep:enforcer:forbid}, $P_a\neq P_b$. Create $\mathcal{P}$ from $\mathcal{\hat{P}}$ as follows: For every $\hat{P}\in \mathcal{\hat{P}}$, include $\hat{P}\setminus S$ in $\mathcal{P}$. It is straight-forward to verify that $G/\mathcal{P}$
is $H$-free. Clearly, \SURP$\leq$\SURPC$\leq k$. Therefore,
$(H_E,v,w)$ is an enforcer for $H$.
\end{proof}

Now we will come up with a reduction from \XFC{H'} to \XFC{H}, where
$H'$ is $H-K$, i.e., $H'$ is isomorphic to $tJ$.

\begin{construction}
    \label{con:unisep}
    Let $G', H_E$ be graphs and $k$ be an integer. 
    Let $v,w$ be two vertices in $H_E$.
    We obtain a graph $G$ from $G'$ by introducing two new
    adjacent vertices $y,z$ and making them adjacent to all 
    vertices of $G'$. Then we obtain a graph $\hat{G}$
    from $G$ as follows: 
    \begin{itemize}
        \item For every edge $xy$ (for $x\in V(G')$), attach 
    $k+1$ copies of $H_E$ by identifying $x$ with $v$ and $y$ with $w$.
    \item For every edge $xz$ (for $x\in V(G')$), attach 
    $k+1$ copies of $H_E$ by identifying $x$ with $v$ and $z$ with $w$.
    \item Attach $k+1$ copies of $H_E$ by identifying $y$ with $v$ and 
    $z$ with $w$.
    \end{itemize} 
\end{construction}

Recall that $H$ is a non-star graph with a universal $K_1$ separator 
or a universal $K_2$ separator, where the universal separator is denoted by $K$.
Let $(G',k)$ be an instance of \XFC{H'}, where $H' = H-K$.
Let $(H_E,v,w)$ be the enforcer we created for $H$ (Lemma~\ref{lem:enforcer}). 
Let $G$ and $\hat{G}$ be the graphs obtained from $(G',H_E,v,w,k)$ by Construction~\ref{con:unisep}.

\begin{lemma}
    \label{lem:unisep2:forward}
    Let $(G',k)$ be a yes-instance of \XFC{H'}. 
    Then $(\hat{G},k)$ is a yes-instance of \XFC{H}.
\end{lemma}
\begin{proof}
To prove that $(\hat{G},k)$ is a yes-instance of \XFC{H}, by
Lemma~\ref{lem:enforcer},
it is sufficient to prove that $(G,k)$ is a yes-instance with a 
solution $\mathcal{P}$ such that $x$ and $y$ are not in the same set in $\mathcal{P}$, $x$ and $z$ are not in the same set in $\mathcal{P}$ (for every $x\in V(G')$), and $y$ and $z$
are not in the same set. 
Let $\mathcal{P}'$ be a partition of vertices of $G'$
such that $G'/\mathcal{P}'$ is $H'$-free and \SURPD$\leq k$.
We obtain $\mathcal{P}$, a partition of vertices of $G$, from $\mathcal{P}'$ by including two singleton sets $P_y=\{y\}$ and 
$P_z=\{z\}$. 
We observe that if $G/\mathcal{P}$ has an induced $H$, then $P_y$
and $P_z$ can act only as universal vertices in the $H$ and hence $G'/\mathcal{P}'$ must contain an induced $H'$, which is a contradiction. It is straight-forward to verify that $\mathcal{P}$ satisfies the rest of the required properties.
\end{proof}

\begin{lemma}
\label{lem:unisep2:backward}
    Let $(\hat{G},k)$ be a yes-instance of \XFC{H}. 
    Then $(G',k)$ is a yes-instance of \XFC{H'}.
\end{lemma}
\begin{proof}
    Let $(\hat{G},k)$ be a yes-instance of \XFC{H}.
    Then Lemma~\ref{lem:enforcer} implies that
    there exists a partition $\mathcal{P}$ of vertices of $G$ such that $P_x\neq P_y$, 
    $P_x\neq P_z$, and $P_y\neq P_z$, for every vertex $x\in V(G')$, where $P_x, P_y, P_z$ denote the sets containing $x,y,z$ respectively.
    Further, $G/\mathcal{P}$ is $H$-free and \SURP$\leq k$.
    This implies that $P_y=\{y\}$ and $P_z=\{z\}$.
    Let $\mathcal{P}'=\mathcal{P}\setminus \{P_y,P_z\}$.
    Clearly, \SURPD$=$\SURP$\leq k$.
    We claim that $G'/\mathcal{P}'$ is $H$-free.
    For a contradiction, assume that $\mathcal{U}\subseteq \mathcal{P}'$
    induces an $H'$ in $G'/\mathcal{P}'$. 
    Then, we obtain that there is an 
     $H$ induced in $G$ by $\mathcal{U}\cup \{P_y\}$, if $|K|=1$, or by 
     $\mathcal{U}\cup \{P_y,P_z\}$, if $|K|=2$. This gives us a contradiction.
\end{proof}

Lemmas~\ref{lem:unisep2:forward} and \ref{lem:unisep2:backward} prove
Lemma~\ref{lem:unisep2}.
\begin{lemma}
    \label{lem:unisep2}
    Let $H$ be a graph with a universal $K_1$ separator or a 
    universal $K_2$ separator, denoted by $K$.
    Assume that $H-K$ is a disjoint union of $t$ copies of a graph $J$,
    for some $t\geq 2$. Let $H$ be not a star and let $H'$ be $tJ$. Then there is a polynomial-time reduction from \XFC{H'} to 
    \XFC{H}.
\end{lemma}
\subsection{Stars}
\label{sub:npc-stars}
Here, we prove that \XFC{H} is \NPC\ whenever $H$ is a star graph of at least 4 vertices. 

\begin{figure}
    \centering
    \begin{tikzpicture}[myv/.style={circle, draw, inner sep=1.5pt}]
    \draw (0,0) ellipse (0.5 and 1);
    \node[blue] at (0,0) {$G'$};
    \node[myv] (x1) at (-1.2,-0.9) {};
    \node[myv] (x2) at (-1.2,-0.6) {};
    \node[myv] (y1) at (-1.2,0.9) {};
    \node[myv] (y2) at (-1.2,0.6) {};
    \draw[dashed, red] (-1.5,1.05) rectangle (-0.9,0.45);
    \draw[dashed, red] (-1.5,-1.05) rectangle (-0.9,-0.45);
    \node[blue] at (-1.2,1.3) {$X$};
    \node[blue] at (-1.2,-1.3) {$Y$};
    \node[blue] at (1.2,-1.3) {$Y'$};
    \node[blue] at (1.2,1.3) {$X'$};

    \node[myv] (zx11) at (-2.2,-1.5) {};
    \node[myv] (zx12) at (-2.2,-1.2) {};
    \draw[dashed, red] (-2.5,-1.05) rectangle (-1.9,-1.65);

    \node[myv] (zx21) at (-2.2,-0.6) {};
    \node[myv] (zx22) at (-2.2,-0.3) {};
    \draw[dashed, red] (-2.5,-0.15) rectangle (-1.9,-0.75);

    \node[myv] (zy11) at (-2.2,1.5) {};
    \node[myv] (zy12) at (-2.2,1.2) {};
    \draw[dashed, red] (-2.5,1.05) rectangle (-1.9,1.65);

    \node[myv] (zy21) at (-2.2,0.6) {};
    \node[myv] (zy22) at (-2.2,0.3) {};
    \draw[dashed, red] (-2.5,0.15) rectangle (-1.9,0.75);

    \node[myv] (xd1) at (1.2,-0.9) {};
    \node[myv] (xd2) at (1.2,-0.6) {};
    \node[myv] (yd1) at (1.2,0.9) {};
    \node[myv] (yd2) at (1.2,0.6) {};
    \draw[dashed, red] (1.5,1.05) rectangle (0.9,0.45);
    \draw[dashed, red] (1.5,-1.05) rectangle (0.9,-0.45);

    \node[myv] (zxd11) at (2.2,-1.5) {};
    \node[myv] (zxd12) at (2.2,-1.2) {};
    \draw[dashed, red] (2.5,-1.05) rectangle (1.9,-1.65);

    \node[myv] (zxd21) at (2.2,-0.6) {};
    \node[myv] (zxd22) at (2.2,-0.3) {};
    \draw[dashed, red] (2.5,-0.15) rectangle (1.9,-0.75);

    \node[myv] (zyd11) at (2.2,1.5) {};
    \node[myv] (zyd12) at (2.2,1.2) {};
    \draw[dashed, red] (2.5,1.05) rectangle (1.9,1.65);

    \node[myv] (zyd21) at (2.2,0.6) {};
    \node[myv] (zyd22) at (2.2,0.3) {};
    \draw[dashed, red] (2.5,0.15) rectangle (1.9,0.75);
    
    \draw (x1) -- (x2);
    \draw (y1) -- (y2);
    \draw (xd1) -- (xd2);
    \draw (yd1) -- (yd2);    
    \draw (zx11) -- (zx12);  
    \draw (zy11) -- (zy12);  
    \draw (zx21) -- (zx22);  
    \draw (zy21) -- (zy22);  
    \draw (zxd11) -- (zxd12);  
    \draw (zyd11) -- (zyd12);  
    \draw (zxd21) -- (zxd22);  
    \draw (zyd21) -- (zyd22);  

    \draw (-1.2,0.45) -- (-1.2,-0.45);
    \draw (1.2,0.45) -- (1.2,-0.45);
    \draw (x1) -- (-1.9,-1.35);
    \draw (x2) -- (-1.9,-0.45);
    \draw (y1) -- (-1.9,1.35);
    \draw (y2) -- (-1.9,0.45);
    \draw (xd1) -- (1.9,-1.35);
    \draw (xd2) -- (1.9,-0.45);
    \draw (yd1) -- (1.9,1.35);
    \draw (yd2) -- (1.9,0.45);
    \draw (-0.9,-0.75) -- (-0.45,-0.4);
    \draw (-0.9,0.75) -- (-0.45,0.4);
    \draw (0.9,-0.75) -- (0.45,-0.4);
    \draw (0.9,0.75) -- (0.45,0.4);

    \end{tikzpicture}
    \caption{Construction of $G$ from $(G',t=3,k=1)$. An edge 
    between two entities represents all possible edges between the 
    vertices of the two entities. This corresponds to the reduction for \XFC{K_{1,3}}.}
    \label{fig:stars}
\end{figure}
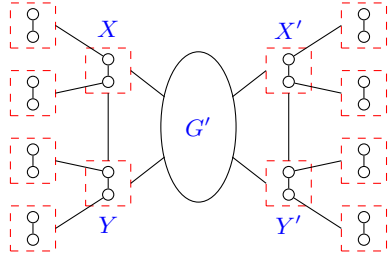

\begin{construction}
\label{con:star}
Let $G'$ be a graph and $k,t$ be integers. We obtain a graph $G$
from $G'$ as follows. 
\begin{itemize}
    \item Introduce four cliques $X,Y,X',Y'$ having $k+1$ vertices each. Make every vertex of $X\cup Y\cup X'\cup Y'$ adjacent to $V(G')$. 
    Similarly, make every vertex of $X$ adjacent to every vertex of $Y$, and 
    every vertex of $X'$ adjacent to every vertex of $Y'$.
    \item For every vertex $z\in X\cup Y\cup X'\cup Y'$,
    introduce $t-2$ cliques $Z_1, Z_2,\ldots, Z_{t-2}$ of $k+1$ vertices each, and make $z$
    adjacent to every vertex of $Z_z = Z_1\cup Z_2\cup \ldots\cup Z_{t-2}$. Let $Z$ denote the union of all $Z_z$s.
\end{itemize}
\end{construction}
An example is given in Figure~\ref{fig:stars}.
For two integers $a,b\geq 1$, by $K_{1,a,b}$ we denote the graph 
obtained by introducing a vertex adjacent to all vertices of $a$
copies of $K_b$s. The following is a trivial observation.

\begin{observation}
\label{obs:star}
Let $(G,k)$ be a yes-instance of \XFC{K_{1,t}} for any integer $t\geq 2$.
Then $G$ does not have $K_{1,t,k+1}$ as an induced subgraph.
\end{observation}

We give a reduction from \XFC{2K_1} to \XFC{K_{1,t}}, for every $t\geq 3$. For an instance $(G',k)$ of \XFC{2K_1}, we obtain an
instance $(G,k)$ of \XFC{K_{1,t}} by applying Construction~\ref{con:star} on $(G',k,t)$.

\begin{lemma}
\label{lem:star:forward}
Let $(G',k)$ be a yes-instance of \XFC{2K_1}. Then $(G,k)$
is a yes-instance of \XFC{K_{1,t}}. 
\end{lemma}
\begin{proof}
Let $\mathcal{P}'$
be a partition of vertices of $G'$ such that \SURPD$\leq k$, $G'/\mathcal{P}'$ is a clique (i.e., $2K_1$-free), and 
every set in $\mathcal{P}'$ induces a connected graph in $G'$.
Construct $\mathcal{P}$ from $\mathcal{P}'$ by introducing singleton
sets corresponding to the vertices $X\cup Y\cup X'\cup Y'\cup Z$.
Clearly, \SURP$=$\SURPD$\leq k$.
It is straight-forward to verify that 
a maximum sized induced star graph in $G/\mathcal{P}$
is $K_{1,t-1}$. Therefore, 
$G/\mathcal{P}$ is $K_{1,t}$-free.
\end{proof}

\begin{lemma}
\label{lem:star:backward}
Let $(G,k)$ be a yes-instance of \XFC{K_{1,t}}.
Then $(G',k)$ is a yes-instance of \XFC{2K_1}.
\end{lemma}
\begin{proof}
Let $\mathcal{P}$ be a partition of vertices of $G$
such that $G/\mathcal{P}$ is $K_{1,t}$-free and \SURP$\leq k$. 
For a vertex $a$ in $G$, let $P_a$ denote the set containing $a$ in $\mathcal{P}$. 
First we claim 
that $P_u\neq P_x$, where $u\in V(G'), x\in X$. If, on the contrary, $P_u=P_x$ for some $u\in V(G')$, we observe that,
in $G/P_u$, there is a $K_{1,t,k+1}$ induced by $P_u$ and the vertices in $Z_x\cup Y\cup X'$. Therefore, by Observation~\ref{obs:star},
the claim is true. By symmetry, we obtain that $P_u\neq P_y$, $P_u\neq P_{x'}$, and $P_u\neq P_{y'}$ for vertices $y\in Y,x'\in X',y'\in Y'$.
This implies that there is no set in $\mathcal{P}$ containing vertices
from both $V(G')$ and $X\cup Y\cup X'\cup Y'\cup Z$.
Therefore, there is a subset $\mathcal{P}'$ of $\mathcal{P}$
such that $\mathcal{P}'$ is a partition of vertices of $G'$ and 
every set in $\mathcal{P}'$ induces a connected subgraph in $G'$.
Now, it is sufficient to prove that that $G'/\mathcal{P}'$ is a clique. 
For a contradiction assume that there is a $2K_1$ induced in $G'/\mathcal{P}'$ by $P_u,P_v\in \mathcal{P}'$.
We observe that for every clique $K$ attached to $x\in X$,
there is a partition $P$ in $\mathcal{P}$ such that $P\subseteq K$ (all the vertices in $K$ cannot be contracted to $x$ as the budget is only $k$ and $K$ has $k+1$ vertices). 
Therefore, for every vertex $x\in X$, $P_x$ is a center of a $K_{1,t-2}$ in $G/\mathcal{P}$, where the leaves are from the $t-2$ cliques attached
to $x$. This along with the sets $P_u, P_v$ induce a $K_{1,t}$
in $G/\mathcal{P}$, which is a contradiction.
\end{proof}

Lemmas~\ref{lem:star:forward} and \ref{lem:star:backward},
and the NP-completeness of \XFC{2K_1} (Proposition~\ref{pro:npc-known})
prove Lemma~\ref{lem:npc:star}.
\begin{lemma}
\label{lem:npc:star}
For $t\geq 3$, \XFC{K_{1,t}} is \NPC.
\end{lemma}
\subsection{Small graphs}
\label{sub:npc-small}
In this section, we prove that \XFC{H} is \NPC\ when $H$
is either a $K_2+K_1$ or a $2K_2$. First we handle the case of $2K_2$ with a simple reduction from \XFC{2K_1}.

\begin{construction}
\label{con:2k2}
Let $G'$ be a graph and $k$ be an integer. We obtain
a graph $G$ by attaching $k+1$ pendant vertices, denoted by a set $Z_u$, to every vertex $u$ in $G'$. Let $Z$ denote the set of all newly
added vertices.
\end{construction}

Let $(G',k)$ be an instance of \XFC{2K_1}. We obtain an instance
$(G,k)$ of \XFC{2K_2} by applying Construction~\ref{con:2k2} on $(G',k)$.

\begin{lemma}
\label{lem:2k2:forward}
Let $(G',k)$ be a yes-instance of \XFC{2K_1}. Then 
$(G,k)$ is a yes-instance of \XFC{2K_2}.
\end{lemma}
\begin{proof}
Let $\mathcal{P}'$ be a partition of vertices of $G'$
such that \SURPD$\leq k$ and $G'/\mathcal{P}'$ is $H$-free.
We obtain a partition $\mathcal{P}$ of vertices of $G$ from $\mathcal{P}'$ by introducing singleton sets corresponding to the
vertices in $Z$. Clearly, \SURP$=$\SURPD$\leq k$.
Since there is no edge induced by the sets corresponds to vertices in $Z$, we obtain that if $G/\mathcal{P}$ is not $2K_2$-free,
then there is an induced $2K_1$ in $G'/\mathcal{P}'$, which is a
contradiction.
\end{proof}

\begin{lemma}
    \label{lem:2k2:backward}
    Let $(G,k)$ be a yes-instance of \XFC{2K_2}.
    Then $(G',k)$ is a yes-instance of \XFC{2K_1}.
\end{lemma}
\begin{proof}
    Let $\mathcal{P}$ be a partition of vertices of $G$
    such that $G/\mathcal{P}$ is $2K_2$-free and \SURP$\leq k$.
    We obtain a partition $\mathcal{P}'$ of vertices of $G'$ as follows:
    For every set $P\in \mathcal{P}$, include $P\setminus Z$ in $\mathcal{P}'$. Since $P$ induces 
    a connected graph in $G$, $P\setminus Z$ induces a connected
    graph in $G'$. Assume that there is a $2K_1$ induced by
    $P'_u, P'_v\in \mathcal{P}'$. Let $u\in P'_u$ and $v\in P'_v$. Since there is a set $Z_u$ of $k+1$ pendant
    vertices attached to $u$ and a set $Z_v$ of $k+1$ pendant vertices
     attached
    to $v$, at least one vertex from $Z_u$ and at least one vertex
    from $Z_v$ form singleton sets in $\mathcal{P}$.
    Then, those two sets along with the sets containing $P'_u$
    and the set containing $P'_v$ in $\mathcal{P}$
    induces a $2K_2$ in $G/\mathcal{P}$, which is a contradiction.
\end{proof}
Now, the NP-completeness of \XFC{2K_2} follows from that of \XFC{2K_1} (Proposition~\ref{pro:npc-known}) and Lemmas~\ref{lem:2k2:forward} and \ref{lem:2k2:backward}.
\begin{lemma}
\label{lem:2k2}
\XFC{2K_2} is \NPC.
\end{lemma}
Next we prove the hardness for \XFC{K_2+K_1} by a reduction 
from \DomaticNum.
Domatic number of a graph is the size of a largest set of disjoint dominating sets of the graph, which partitions the vertices of the graph. For example, the domatic number of a 
complete graph is the number of vertices of it, and that of a star graph is 2. The \DomaticNum\ problem is to find whether the domatic number of the input graph is at least $k$ or not.
It is known that \DomaticNum\ is \NPC~\cite{DBLP:books/fm/GareyJ79} even for various classes of graphs~\cite{DBLP:journals/dam/Bonuccelli85,DBLP:journals/ipl/KaplanS94}. 
Recall that $K_2+K_1$-free graphs are exactly the class of 
complete multipartite graphs. The reduction that we use is 
exactly the same as the reduction for the NP-completeness of 
Hadwiger number problem (which is equivalent to \XFC{2K_1}) 
described by Eppstein~\cite{DBLP:journals/jgaa/Eppstein09}. 
The proof requires some adaptation. 

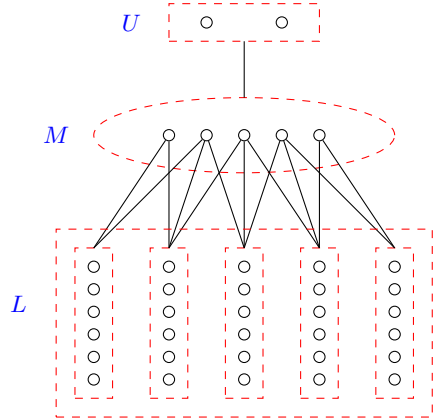
\begin{figure}
    \centering
\begin{tikzpicture}[myv/.style={circle, draw, inner sep=1.5pt}]
     \draw[dashed, red] (-1, -0.25) rectangle (1,0.25);
     \node[myv] (u1) at (-0.5,0) {};
     \node[myv] (u2) at (0.5,0) {};
    \node[blue] at (-1.5,0) {$U$};

      \draw [dashed, red] (0, -1.5) ellipse (2 and 0.5);
     \node[myv] (m3) at (0,-1.5) {};
     \node[myv] (m2) at (-0.5,-1.5) {};
     \node[myv] (m1) at (-1,-1.5) {};
     \node[myv] (m4) at (0.5,-1.5) {};
     \node[myv] (m5) at (1,-1.5) {};
    \node[blue] at (-2.5,-1.5) {$M$};

         \draw (0,-0.25) -- (0,-1);    

     \draw[dashed, red] (-0.25, -3) rectangle (0.25,-5);
     \node[myv] (l31) at (0,-3.25) {};
     \node[myv] (l32) at (0,-3.55) {};
     \node[myv] (l33) at (0,-3.85) {};
     \node[myv] (l34) at (0,-4.15) {};
     \node[myv] (l35) at (0,-4.45) {};
     \node[myv] (l36) at (0,-4.75) {};
         \draw (m3) -- (0,-3);    
         \draw (m4) -- (0,-3);         
         \draw (m2) -- (0,-3);

\begin{scope}[shift={(-1,0)}]
     \draw[dashed, red] (-0.25, -3) rectangle (0.25,-5);
     \node[myv] (l21) at (0,-3.25) {};
     \node[myv] (l22) at (0,-3.55) {};
     \node[myv] (l23) at (0,-3.85) {};
     \node[myv] (l24) at (0,-4.15) {};
     \node[myv] (l25) at (0,-4.45) {};
     \node[myv] (l26) at (0,-4.75) {};
         \draw (m3) -- (0,-3);    
         \draw (m2) -- (0,-3);
        \draw (m1) -- (0,-3);

\end{scope}

\begin{scope}[shift={(-2,0)}]
     \draw[dashed, red] (-0.25, -3) rectangle (0.25,-5);
     \node[myv] (l11) at (0,-3.25) {};
     \node[myv] (l12) at (0,-3.55) {};
     \node[myv] (l13) at (0,-3.85) {};
     \node[myv] (l14) at (0,-4.15) {};
     \node[myv] (l15) at (0,-4.45) {};
     \node[myv] (l16) at (0,-4.75) {};
         \draw (m1) -- (0,-3);    
         \draw (m2) -- (0,-3);
\end{scope}

\begin{scope}[shift={(1,0)}]
     \draw[dashed, red] (-0.25, -3) rectangle (0.25,-5);
     \node[myv] (l41) at (0,-3.25) {};
     \node[myv] (l42) at (0,-3.55) {};
     \node[myv] (l43) at (0,-3.85) {};
     \node[myv] (l44) at (0,-4.15) {};
     \node[myv] (l45) at (0,-4.45) {};
     \node[myv] (l46) at (0,-4.75) {};
         \draw (m5) -- (0,-3);    
         \draw (m4) -- (0,-3);
         \draw (m3) -- (0,-3);

\end{scope}

\begin{scope}[shift={(2,0)}]
     \draw[dashed, red] (-0.25, -3) rectangle (0.25,-5);
     \node[myv] (l51) at (0,-3.25) {};
     \node[myv] (l52) at (0,-3.55) {};
     \node[myv] (l53) at (0,-3.85) {};
     \node[myv] (l54) at (0,-4.15) {};
     \node[myv] (l55) at (0,-4.45) {};
     \node[myv] (l56) at (0,-4.75) {};
         \draw (m5) -- (0,-3);
         \draw (m4) -- (0,-3);

\end{scope}

     \draw[dashed, red] (-2.5, -2.75) rectangle (2.5,-5.25);
     \node[blue] at (-3,-3.75) {$L$};



    


    \end{tikzpicture}
    \caption{Construction of instance $(G,h)$ of \XFC{K_2+K_1} from an instance $(P_5,d=2)$ of \DomaticNum. Here $h=n(n+1)+d=5*6+2=32$. Ellipses represent independent sets, rectangles represent cliques, and an edge between two entities denotes all possible edges between the vertices corresponding to those entities.}
    \label{fig:k2uk1}
\end{figure}

Let $(G',d)$ be an instance of \DomaticNum\ such that $G'$ has 
$n\geq 2$ vertices. 
Let $\{v_1,v_2,\ldots,v_n\}$ be the vertices of $G'$.
We can safely assume that there are no universal vertices in $G'$ (otherwise, remove it and decrease $d$ by 1 to obtain an equivalent instance). We obtain a graph $G$ as follows:
\begin{itemize}
    \item Introduce a clique $U = \{u_1, u_2, \ldots, u_d\}$ of size $d$, called the upper vertices.
    \item Introduce an independent set $M = \{m_1, m_2, \ldots, m_n\}$ of size $n$, called the middle vertices.
    \item Introduce a clique $L = L_1\cup L_2\cup \ldots\cup L_n$ of $n(n+1)$ vertices, where $L_i = \{\ell_{i,1}, \ell_{i,2}, \ldots, \ell_{i,n+1}\}$ is a clique of $n+1$ vertices, for $1\leq i\leq n$. These vertices are called the lower vertices.
    \item Make every vertex in $U$ adjacent to every vertex in $M$.
    \item The vertex $m_i$ is adjacent to (every vertex of) $L_j$ if and only if $v_i$ dominates $v_j$, i.e., either $i=j$ or $v_i$ is adjacent to $v_j$ in $G'$.
\end{itemize}

Let $N$ be the number of vertices in $G$, i.e., $N=n(n+1)+n+d$. Let $h = n(n+1)+d$. This completes the reduction. An example is given in Figure~\ref{fig:k2uk1}.

The idea behind 
Eppstein's reduction is the following: If $G'$ has $d$ 
disjoint dominating sets, then a partition of vertices 
of $G$ can be obtained by the subsets of $M$ corresponding to 
the dominating sets of $G'$ (subsets of $M$ is an independent set, so take
one vertex from $U$ to make sure that each set induces a 
connected subgraph). On the other hand, if $(G,h)$ is 
a yes-instance of \XFC{2K_1}, then there exists a partition 
$\mathcal{P}$ (which is a solution) such that each subset of 
$M$ contained in $P_u$ (the set containing a vertex $u\in U$ in 
the partition) implies a dominating set in $G'$. It turns out that, 
the largest complete multipartite graph that can be obtained from 
$G$ by contractions is a complete graph.

\begin{lemma}
\label{lem:k2k1:forward}
If the domatic number of $G'$ is at least $d$, then there is a partition $\mathcal{P}$ of vertices of $G$ such that every set in $\mathcal{P}$ induces a connected graph in $G$ and $G/\mathcal{P}$ 
is a complete multipartite graph of at least $h$ vertices.
\end{lemma}
\begin{proof}
Assume that $\{D_1, D_2, \ldots, D_d\}$ is a set of $d$ mutually disjoint dominating sets of $G'$. We define a partition $\mathcal{P}$
of vertices of $G$ as follows:
\begin{itemize}
    \item Let $M_i$ be the subset of vertices of $M$ corresponding to $D_i$, i.e., $M_i=\{m_j:v_j\in D_i\}$. Form a set $M_i\cup \{u_i\}$, for $1\leq i\leq d$.
    \item Every vertex in $L$ forms a singleton set in $\mathcal{P}$.
\end{itemize}
It is straight-forward to verify that there are $h$ sets in $\mathcal{P}$, each set induces a connected graph in $G$, and $G/\mathcal{P}$ is a complete graph.
\end{proof}

Let $\mathcal{P}$ be a partition of vertices of $G$ such that each
set in $\mathcal{P}$ forms a connected subgraph and $G/\mathcal{P}$
is a complete multipartite graph. Further assume that $|\mathcal{P}| \geq h$.
Observations~\ref{obs:k2k1:1}, \ref{obs:k2k1:2}, \ref{obs:k2k1:3}, and
Lemma~\ref{lem:k2k1:backward:clique} are essentially taken from \cite{DBLP:journals/jgaa/Eppstein09}.
Observation~\ref{obs:k2k1:1} follows from the fact that
\SURP$+|\mathcal{P}|=N$.
\begin{observation}
    \label{obs:k2k1:1}
    \SURP\ $\leq n$.
\end{observation}

\begin{observation}
\label{obs:k2k1:2}
For each $1\leq i\leq n$, there is a set $L_i'\in \mathcal{P}$ such
that $L_i'\subseteq L_i$.
\end{observation}
\begin{proof}
    Assume that for every vertex $\ell_{i,j}$ in $L_i$, 
    the set in $\mathcal{P}$ containing $\ell_{i,j}$ contains at least one vertex from outside $L_i$. Then \SURP\ is at least $n+1$ as there are $n+1$
    vertices in $L_i$, which contradicts Observation~\ref{obs:k2k1:1}.
\end{proof}

Observations \ref{obs:k2k1:3}, \ref{obs:k2k1:4},
and Lemma~\ref{lem:k2k1:backward:clique} assumes that $G/\mathcal{P}$
is a clique.

\begin{observation}
    \label{obs:k2k1:3}
    $\mathcal{P}$ cannot contain a singleton subset of $M$.
\end{observation}
\begin{proof}
    Since there are no universal vertices in $G'$, the degree of a 
    vertex in $M$ is at most $(n-1)(n+1)+d=n(n+1)+d-(n+1)<h-1$.
    Since $G/\mathcal{P}$ is a clique of size at least $h$, we obtain that
    $\mathcal{P}$ cannot contain a singleton subset of $M$.
\end{proof}

Observation~\ref{obs:k2k1:4} follows from Observation~\ref{obs:k2k1:2}
and the assumption that $G/\mathcal{P}$ is a clique.
\begin{observation}
    \label{obs:k2k1:4}
    If $P\in \mathcal{P}$ contains a vertex in $U$,
    then it contains a vertex from $M$.
\end{observation}

\begin{lemma}
    \label{lem:k2k1:backward:clique}
    $G'$ has a set of at least $d$ mutually disjoint dominating sets.
\end{lemma}

\begin{proof}
By Observation~\ref{obs:k2k1:1}, \SURP\ $\leq n$. Let $\mathcal{P}_M$ be the subset of $\mathcal{P}$ such that each set in $\mathcal{P}_M$
contains at least one vertex from $M$. By Observation~\ref{obs:k2k1:3}, each set $P\in \mathcal{P}_M$ contains at least two vertices. Since $M$ is an independent set
and $P$ induces a connected subgraph of $G$, we obtain that $P$
is not a subset of $M$. Therefore, cost($\mathcal{P}_M$)$= n$.
This implies that $P$ contains exactly one vertex outside $M$. 
By Observation~\ref{obs:k2k1:4}, every set $P_i$ containing a vertex
$u_i$ has at least one vertex from $M$. As a set containing a vertex
from $M$ contains exactly one vertex outside $M$, we obtain that
$P_i$ does not contain any vertex from $L$. Now we obtain $d$
mutually disjoint dominating sets of $G'$ as follows. 
For every $P_i$, $D_i$ is the vertices in $G'$ corresponding to $P_i\cap M$, i.e., $D_i=\{v_j|m_j\in P_i\}$. It is sufficient to 
prove that $D_i$ is a dominating set of $G'$. Assume that a vertex $v_\ell$ is not dominated by $D_i$ in $G'$. By Observation~\ref{obs:k2k1:2}, there is a set $L_i'\in \mathcal{P}$ such that $L_i'\subseteq L_i$. Then, $P_i$ is not adjacent to that set, which is a 
contradiction.
\end{proof}

\begin{lemma}
    \label{lem:k2k1:main}
    Assume that there is a partition $\mathcal{P}$ of vertices of $G$ into sets inducing connected subgraphs such that $G/\mathcal{P}$ is a complete multipartite graph $Q$ on at least $h$ vertices. Then $Q$ is a clique. 
\end{lemma}
\begin{proof}
    For a contradiction, assume that $Q$ is not a clique. Let $r, s$ be two nonadjacent vertices of $Q$. Let $R, S$ be the sets in $\mathcal{P}$ containing $r,s$ respectively.

    Claim 1: $R$ cannot be a subset of $U$.

    Proof of Claim 1: Assume that $R\subseteq U$. Clearly, $S$
    cannot contain any vertex from $U\cup M$. Therefore, $S\subseteq L$.
    Without loss of generality, assume that $S$ contains at least one vertex, say $l_{i,j}$, from $L_i$. Then every other vertex in $L_i$
    contributes towards the cost of $\mathcal{P}$. 
    For example, if $l_{i,q}$ ($q\neq j$) is part of $S$, then it contributes toward the cost of $\mathcal{P}$. If $l_{i,q}$ is not
    part of $S$, then the set containing $l_{i,q}$ must contain some other vertices from outside $L_i$ to make sure that it is adjacent to $R$ 
    (otherwise there is a $K_1+K_2$ induced by $R, S,$ and the set containing $l_{i,q}$). 
    Since \SURP\ is at most $n$ (Observation~\ref{obs:k2k1:1}), 
    this implies that sets in $\mathcal{P}$
    containing vertices from $L_i\setminus \{\ell_{i,j}\}$
    are the only sets containing at least two vertices and 
    all other sets are singleton sets. Further, each set $P\in \mathcal{P}$
    containing at least one vertex from $L_i\setminus \{\ell_{i,j}\}$
    contains exactly one
    vertex outside $L_i$ and that vertex must be from $M$ (to make sure that $P$ is adjacent to $R$ - otherwise there is a $K_2+K_1$
    induced by $R,S$, and $P$). This implies that every vertex in
    $L\setminus L_i$ forms a singleton set in $\mathcal{P}$. Then
    any such set (recall that $n\geq 2$) along with $R$ and $S$ induces a $K_2+K_1$,
    which is a contradiction.

    Claim 2: It cannot happen that $R$ contains a middle vertex and $S$ contains a lower vertex.

    Proof of Claim 2: For a contradiction, assume that $R$ contains a middle vertex $m$ and $S$ contains a lower vertex $l_{i,j}\in L_{i}$. Then every other vertex in $L_i$
    contributes toward the cost of $\mathcal{P}$ (as obtained in the proof of Claim 1). Therefore,
    $L_i$ contributes $n$ toward the \SURP.
    This implies that the sets $P\in \mathcal{P}$ containing 
    at least one vertex from $L_i\setminus \{\ell_{i,j}\}$
    are the only sets containing at least two vertices
    and all other sets in $\mathcal{P}$ are singleton sets.
    Further, each such set $P$ can contain only one vertex outside $L_i$ - clearly, that extra vertex cannot be from $U$. 
    Then $S$ is a singleton set and every vertex in $U$ forms a
    singleton set in $\mathcal{P}$. Then there is a $K_2+K_1$
    induced in $Q$ by $R,S,$ and one singleton set formed by a vertex in $U$, which is a contradiction.

    Claim 1 and Claim 2 implies that both $R$ and $S$ are subsets of $M$. Since both $R$ and $S$ must induce connected subgraph of $G$,
    we obtain that both $R$ and $S$ are singleton sets. Let $R=\{m\}$.
    Let $L_i$ be nonadjacent to $m$ (this is guaranteed as there are
    no universal vertices in $G'$). By Claim 2, a set $P\in \mathcal{P}$ containing 
    at least one vertex from $L_i$ must be adjacent to $m$.
    Then every vertex in $L_i$
    contribute toward \SURP. This is a contradiction as $|L_i|=n+1$
    and \SURP\ $\leq n$.   
\end{proof}

Lemmas~\ref{lem:k2k1:forward}, \ref{lem:k2k1:backward:clique}, \ref{lem:k2k1:main} imply that there is a polynomial-time reduction from 
\DomaticNum\ to \XFC{K_2+K_1}.

\begin{lemma}
\label{lem:k2k1}
\XFC{K_2+K_1} is \NPC.
\end{lemma}
\subsection{Putting them together}
Recall that the reduction from \VC\ in Section~\ref{sub:npc-general}
does not handle disconnected graph. This is the main ingredient that remains to be added to obtain the main result of the section. This turns out to be easy.
Guo~\cite{DBLP:phd/ndltd/Guo13} has a reduction for transferring
the hardness of \XFC{H'} to \XFC{H}, where $H'$ is any component of $H$.
\label{sub:npc-ptt}
\begin{proposition}[\cite{DBLP:phd/ndltd/Guo13}]
\label{pro:disconnected}
Let $H$ be a disconnected graph. Let $H'$ be any component of it.
Then there is a polynomial-time reduction from \XFC{H'} to \XFC{H}.
\end{proposition}

Proposition~\ref{pro:disconnected} does not help us to prove the 
hardness when every component of $H$ is either a $K_1$ or a $K_2$. 
But there are simple reductions to handle them.

\begin{lemma}
\label{lem:disconnected:k1}
Let $H$ be a disconnected graph with an isolated vertex $v$.
There is a polynomial-time reduction from \XFC{(H-v)} to \XFC{H}.
\end{lemma}
\begin{proof}
It is straight-forward to verify that $(G',k)$ is a yes-instance 
of \XFC{(H-v)} if and only if $(G'+K_1,k)$ is a yes-instance of 
\XFC{H}.
\end{proof}

Repeated application of Lemma~\ref{lem:disconnected:k1}
implies that there is a polynomial-time reduction from \XFC{2K_1} 
to \XFC{tK_1}, for every $t\geq 3$. Then the NP-Completeness of 
\XFC{2K_1} (Proposition~\ref{pro:npc-known}) implies Lemma~\ref{lem:npc:tk1}.
\begin{lemma}
    \label{lem:npc:tk1}
    For every $t\geq 3$, \XFC{tK_1} is \NPC.
\end{lemma}

Now, we handle the case when $H$ is a disjoint union of $t$ copies of $K_2$.
\begin{lemma}
    \label{lem:disconnected:k2}
    Let $H=tK_2$, for any integer $t\geq 3$ and let $H'$ be $(t-1)K_2$.
    There is a polynomial-time reduction from \XFC{H'}
    to \XFC{H}.
\end{lemma}
\begin{proof}
    It is straight-forward to verify that $(G',k)$ is a 
    yes-instance of \XFC{H'} if and only if $(G'+K_{k+2},k)$
    is a yes-instance of \XFC{H}.
\end{proof}

Repeated application of Lemma~\ref{lem:disconnected:k2} 
and the NP-completeness of \XFC{2K_2} (Lemma~\ref{lem:2k2})
give us 
the following Lemma.
\begin{lemma}
    \label{lem:npc:tk2}
    For every $t\geq 3$, \XFC{tK_2} is \NPC. 
\end{lemma}
Now we are ready to prove the main result of this section.
\begin{theorem}
\label{thm:npc}
Let $H$ be any graph other than $K_1$ and $K_2$.
Then \XFC{H} is \NPC.
\end{theorem}
\begin{proof}
    We prove this by induction on $n$, the number of vertices of $H$.
    The base cases are when $n=2$ and $n=3$, i.e., when $H$
    is $2K_1$ or $P_3$ or triangle (Proposition~\ref{pro:npc-known}), or 
    $3K_1$ (Lemma~\ref{lem:npc:tk1}), or $K_2+K_1$ (Lemma~\ref{lem:k2k1}). Assume that $n\geq 4$.
    
    Let $H$ be a disconnected graph. Assume that $H$ has a component $H'$
    with at least three vertices. By Proposition~\ref{pro:disconnected},
    there is a polynomial-time reduction from \XFC{H'} to \XFC{H}.
    Then we are done by induction hypothesis. Assume that every
    component of $H$ is either a $K_2$ or a $K_1$. If $H$ has 
    an isolated vertex, then we are done by Lemma~\ref{lem:disconnected:k1}. If there are no isolated 
    vertex in $H$, then $H$ is isomorphic to $tK_2$, for $t\geq 2$.
    Then we are done by Lemma~\ref{lem:npc:tk2}. 

    Let $H$ be a connected graph. If $H$ is complete, then 
    Proposition~\ref{pro:npc-known} is sufficient.
    Assume that $H$ is non-complete. Therefore, there is a 
    non-universal vertex in $H$. Assume that $H$ has neither
    a universal $K_1$ separator nor a universal $K_2$ separator.
    Then we are done by Lemma~\ref{lem:npc:general}.
    Assume that $H$ has either a universal $K_1$ separator or a 
    universal $K_2$ separator, denoted by $K$.
    Further assume that $H$ is not a star.
    Let $H-K$ has at least two non-isomorphic components.
    Let $J$ be any component in $H-K$ with least number of vertices.
    Let $H'$ be obtained from $H$ by removing all copies of $J$
    in $H-K$. Then by Lemma~\ref{lem:unisep1}, there is a
    polynomial-time reduction from \XFC{H'} to \XFC{H}, and we 
    are done. Assume that $H-K$ is disjoint union of $t$ copies of 
    a graph $J$, for some $t\geq 2$. Then Lemma~\ref{lem:unisep2}
    gives us a polynomial-time reduction from \XFC{tJ} to \XFC{H}.
    For the last case, assume that $H$ is a star graph. Then
    we are done by Lemma~\ref{lem:npc:star}.
\end{proof}
\section{W[2]-hardness}
\label{sec:whardness}

In this section, we prove that \XFC{H} is W[2]-hard whenever $H$
is a tree, except for 7 trees. All our reductions are from \DS, which
is well-known to be a W[2]-hard problem. 
First we obtain a reduction for all trees which are neither stars nor bistars. Then we come up with a reduction for a subset of bistars, a corner 
case of the same proves the case of stars. Then we come up with a 
reduction which handles the remaining bistars.

Recall that a dominating set
$D$ of a graph $G$ is a subset of vertices of $G$ such that every vertex of $G$ is either in $D$ or adjacent to a vertex in $D$.
The objective of the \DS\ problem is to check whether a graph has a dominating
set of size at most $k$ or not.
\subsection{A general reduction for trees}
\begin{figure}
  \centering
  \begin{subfigure}[b]{0.3\textwidth}
    \centering
    \begin{tikzpicture}[myv/.style={circle, draw, inner sep=1.5pt}]
    \node[myv,label={[blue]10:$w$}] (w) at (0,0) {};
    
    \node[myv] (w21) at (-0.5,-1.5) {};
    \node[myv] (w22) at (-0.5,-2.5) {};
    \node[myv] (w23) at (0,-1.5) {};
    \node[myv] (w24) at (0.5,-1.5) {};
    \node[myv] (w25) at (0.5,-2.5) {};

    \draw (w) -- (w21);
    \draw (w) -- (w23);
    \draw (w) -- (w24);
    \draw (w21) -- (w22);
    \draw (w24) -- (w25);

    
    \end{tikzpicture}
    \caption{$H$}
    \label{fig:wgeneral-H}
  \end{subfigure}%
  \begin{subfigure}[b]{0.5\textwidth}
    \centering
    \begin{tikzpicture}[myv/.style={circle, draw, inner sep=1.5pt}]
    \node[myv,label={[blue]10:$w$}] (w) at (0,0) {};
    \node[myv,label={[blue]90:$x_2$}] (x2) at (0,1) {};
    \node[myv,label={[blue]90:$x_1$}] (x1) at (-1,1) {};
    \node[myv,label={[blue]90:$x_3$}] (x3) at (1,1) {};
    \draw[dashed, red] (-1.25,1.75) rectangle (1.25,0.75);

    \draw (x1) -- (x2);
    \draw (x2) -- (x3);
    \draw (w) -- (x1);
    \draw (w) -- (x2);
    \draw (w) -- (x3);
    \draw[bend left=50] (x1) to (x3);

    \node[myv] (w11) at (-2.5,0) {};
    \node[myv] (w12) at (-1.5,0) {};
    \node[myv] (w13) at (-1.5,-0.5) {};
    \node[myv] (w14) at (-2.5,-1) {};
    \node[myv] (w15) at (-1.5,-1) {};

    \node[myv] (w31) at (2.5,0) {};
    \node[myv] (w32) at (1.5,0) {};
    \node[myv] (w33) at (1.5,-0.5) {};
    \node[myv] (w34) at (2.5,-1) {};
    \node[myv] (w35) at (1.5,-1) {};

    \draw (w11) -- (w12);
    \draw (w14) -- (w15);    
    \draw (w) -- (w12);
    \draw (w) -- (w15);
    \draw (w) -- (w13);

    \draw (w31) -- (w32);
    \draw (w34) -- (w35);    
    \draw (w) -- (w32);
    \draw (w) -- (w35);
    \draw (w) -- (w33);

    \node[circle,draw,dashed,red, minimum size=1.75cm,label={[blue]90:$W_1$}] (c1) at  (-2,-0.5) {};    
    \node[circle,draw,dashed,red, minimum size=1.75cm,label={[blue]90:$W_3$}] (c3) at  (2,-0.5) {};
    \node[circle,draw,dashed,red, minimum size=1.75cm,label={[blue]0:$W_2$}] (c2) at  (0,-2) {};
    
    \node[myv] (w21) at (-0.5,-1.5) {};
    \node[myv] (w22) at (-0.5,-2.5) {};
    \node[myv] (w23) at (0,-1.5) {};
    \node[myv] (w24) at (0.5,-1.5) {};
    \node[myv] (w25) at (0.5,-2.5) {};

    \draw (w) -- (w21);
    \draw (w) -- (w23);
    \draw (w) -- (w24);
    \draw (w21) -- (w22);
    \draw (w24) -- (w25);

    \draw[very thick] (x1) -- (c1);
    \draw[very thick] (x1) -- (c2);
    \draw[very thick] (x2) -- (c1);
    \draw[bend left=50,very thick] (x2) to (c2);
    \draw[very thick] (x2) -- (c3);
    \draw[very thick] (x3) -- (c2);
    \draw[very thick] (x3) -- (c3);

    
    \end{tikzpicture}
    \caption{$G$}
    \label{fig:wgeneral-G}
  \end{subfigure}%
  \caption{Construction of $G$ from $(G'=P_3,H)$ by Construction~\ref{con:wgeneral}}
  \label{fig:wgeneral-all}
\end{figure}
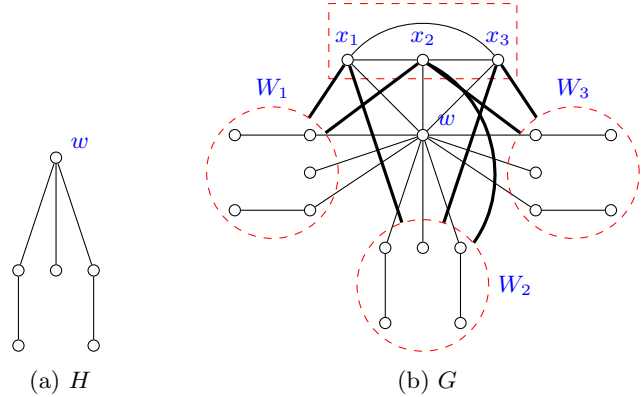

Here we handle all trees which are neither stars nor bistars.
The reduction that we use 
is an adapted version of a reduction used in \cite{DBLP:conf/iwpec/CaiG13} (to handle 3-connected graphs) and a reduction used in \cite{DBLP:conf/iwpec/LokshtanovMS13} (to handle cycles).
\begin{construction}
\label{con:wgeneral}
Let $G', H$ be graphs and $k$ be an integer. Let $\{v_1,v_2,\ldots,v_n\}$ be the set of vertices of $G'$. We construct
a graph $G$ from $(G',H,k)$ as follows.

\begin{itemize}
    \item Introduce a clique $X=\{x_1,x_2,\ldots,x_n\}$.
    \item Introduce $n$ copies of $H$ denoted by $H_1,H_2,\ldots,H_n$. Let $V_i$ denote the set of vertices of $H_i$, for $1\leq i\leq n$.
    \item Let $w$ be any vertex in $H$. Identify $w$'s of all copies of 
    $H$. Let the vertex obtained so be denoted by $w$. Let the remaining vertices in each copy $H_i$ be denoted by $W_i$, i.e.,
    $W_i\cup \{w\}$ induces $H$, for $1\leq i\leq n$.
    \item Make $w$ adjacent to every vertex in $X$.
    \item Make $x_i$ adjacent to every vertex in $W_j$ if and only
    if $i=j$ or $v_i$ is adjacent to $v_j$ in $G'$.
\end{itemize}
This completes the construction. An example is shown in Figure~\ref{fig:wgeneral-all}.
\end{construction}

Let $(G',k)$ be an instance of \DS. Let $H$ be a tree which
is neither a star nor a bistar. We obtain a graph $G$ from $(G',H,k)$
by applying Construction~\ref{con:wgeneral}.

\begin{lemma}
\label{lem:wgeneral:forward}
Let $(G',k)$ be a yes-instance of \DS. 
Then $(G,k)$ is a yes-instance of \XFC{H}.
\end{lemma}
\begin{proof}
Let $D$ be a dominating set of size at most $k$ of $G'$.
Let $F=\{wx_i| v_i\in D\}$. Clearly, $|F|=|D|\leq k$.
We claim that $G/F$ is $H$-free.
Let $w$ itself denote the vertex obtained by contracting the edges in 
$F$. 
To get a contradiction, assume that there is an $H$ induced by a set
$U$ of vertices of $G/F$. We observe that $w$ is a universal vertex in $G/F$,
due to the fact that $D$ is a dominating set of $G'$.
Since $H$ does not contain a universal vertex (a tree has a universal vertex if and only if it is a star), $w$ cannot be in $U$.
Since $H$ is triangle-free, $U$ can have at most two vertices from $X$.
Assume that $U$ has no vertex in $X$. Then $U$ must be a subset of $W_i$, which is a contradiction (observe that $W_i$ and $W_j$ are nonadjacent in $G/F$ for $i\neq j$). 
Assume that $U$ has exactly one vertex, say $x_i$, from $X$.
Then the rest of the vertices are from $W_j$s adjacent to $x_i$.
Recall that $x_i$ is adjacent to either all or none of vertices in $W_j$. Since $H$ is triangle-free, the vertices in $U$ from $W_j$s
adjacent to $x_i$ form an independent set. Then $H$ is a star, which
is a contradiction. If $U$ has exactly two vertices from $X$, then with similar arguments, we obtain that $H$ is a bistar, which is a 
contradiction.
\end{proof}

\begin{lemma}
    \label{lem:wgeneral:backward}
    Let $(G,k)$ be a yes-instance of \XFC{H}.
    Then $(G',k)$ is a yes-instance of \DS.
\end{lemma}
\begin{proof}
    Let $F'$ be a subset of edges of $G$ such that
    $G/F'$ is $H$-free and $|F'|\leq k$.
    We construct a new solution $F$ from $F'$ as follows.
    If $F'$ contains an edge touching $W_j$, then we replace that
    edge with the edge $wx_j$ in $F$.
    Clearly $|F|\leq |F'|\leq k$. 
    Since an edge touching $W_j$ kills only the $H$ induced by
    $W_j\cup \{w\}$, which is killed by $wx_j$, we obtain that $G/F$
    is $H$-free. Let $\mathcal{P}$ be the partition of vertices
    of $G$ corresponds to $F$. Let $P_w$ be the set in $\mathcal{P}$ containing $w$. Let $D = P_w\cap X$. Clearly, $|D|\leq |$\SURP$|\leq k$. We claim that $D' = \{v_i|x_i\in D\}$ is a dominating set
    of $G'$. Assume that there is a vertex $v_j$ in $G'$ not dominated
    by $D'$. Then $W_j\cup \{w\}$ induces an $H$ in $G/F$, which
    is a contradiction.
\end{proof}

Lemmas~\ref{lem:wgeneral:backward} and
\ref{lem:wgeneral:forward} imply that there 
is a parameterized reduction from \DS\ to 
\XFC{H}.

\begin{lemma}
\label{lem:wgeneral}
Let $H$ be a tree which is neither a star nor a bistar.
Then \XFC{H} is W[2]-hard.
\end{lemma}
\subsection{Stars and Bistars}
First we generalize a reduction given in \cite{DBLP:phd/ndltd/Guo13} for \XFC{K_{1,4}}. This generalized reduction covers all bistars $T_{t,t'}$ such that $t\geq 3$ and $t> t'\geq 0$. As a boundary case (when $t'=0$) we obtain hardness result for all stars of at least 5 vertices. 
\begin{construction}
\label{con:wbistars1}
Let $G'$ be a graph with a vertex set $V'=\{v_1,v_2,\ldots,v_n\}$ and let $k,t,t'$ be integers. We construct a graph $G$ from $(G',k,t,t')$
as follows.
\begin{itemize}
    \item Create two cliques $X=\{x_1,x_2,\ldots,x_n\}, Y=\{y_1,y_2,\ldots,y_n\}$.
    \item Make $x_i$ adjacent to $y_j$ if and only if $i=j$ or
    $v_iv_j$ is an edge in $G'$.
    \item Create a vertex $w$ and two cliques $A$ and $B$ of $k+1$ vertices each.
    \item Make $w$ adjacent to all vertices of $X\cup A\cup B$.
    Make $A\cup B\cup Y$ a clique.
    \item Introduce $t-1$ cliques with $k+1$ vertices each and make them
    adjacent to $A$. Let $A'$ denote the set of these vertices.
    \item Introduce $t-1$ cliques with $k+1$ vertices each and make
    them adjacent to $B$. Let $B'$ denote the set of these vertices..
    \item Introduce $t-1$ cliques with $k+1$ vertices each and make 
    them adjacent to $X$. Let $X'$ denote the set of these vertices.
    \item Introduce a clique of $k+1$ vertices and make
    it adjacent to $Y$. Let $Y'$ denote the set of these vertices.
    \item     For every vertex in $A'\cup B'\cup X'\cup Y'$,
    attach $t'$ degree-1 vertices.
\end{itemize}
\end{construction}

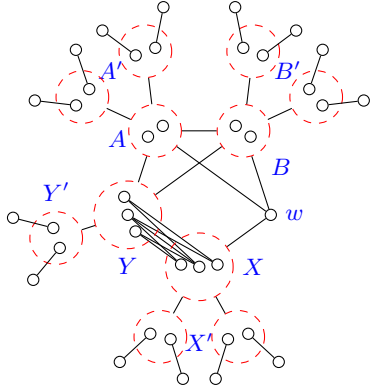
\begin{figure}
    \centering
\begin{tikzpicture}[myv/.style={circle, draw, inner sep=1.5pt}]

    \node (o) at (0,0) {};
    \node[myv,label={[blue]0:$w$}] (w) at (342:1) {};
    \node[myv] (a1) at (133:1) {};
    \node[myv] (a2) at (119:1) {};
    \node[myv] (b1) at (61:1) {};
    \node[myv] (b2) at (47:1) {};
    \node[myv] (y2) at (198:1) {};
    \node[myv] (y1) at (184:1) {};
    \node[myv] (y3) at (212:1) {};
    \node[myv] (x2) at (270:1) {};
    \node[myv] (x1) at (256:1) {};
    \node[myv] (x3) at (284:1) {};

    \draw (x1) -- (y3);
    \draw (x1) -- (y2);
    \draw (x2) -- (y1);
    \draw (x2) -- (y2);
    \draw (x2) -- (y3);
    \draw (x3) -- (y2);
    \draw (x3) -- (y1);

    \node[blue] at (147:1.3) {$A$};
    \node[blue] at (126:2) {$A'$};
    \node[blue] at (54:2) {$B'$};
    \node[blue] at (180:1.9) {$Y'$};
    \node[blue] at (270:2) {$X'$};

    \node[circle,dashed,red,minimum size=0.7cm,draw] (ra) at (126:1) {};
    \node[circle,dashed,red,minimum size=0.7cm,draw,label={[blue]-45:$B$}] (rb) at (54:1) {};
    \node[circle,dashed,red,minimum size=0.9cm,draw,label={[blue]0:$X$}] (rx) at (270:1) {};
    \node[circle,dashed,red,minimum size=0.9cm,draw,label={[blue]270:$Y$}] (ry) at (198:1) {};

    \draw (ra) -- (rb);
    \draw (ra) -- (ry);
    \draw (rb) -- (ry);
    \draw (w) -- (rx);
    \draw (w) -- (ra);
    \draw (w) -- (rb);

    \node[myv] (a11) at (145:2) {};
    \node[myv] (a12) at (137:2) {};

    \node[myv] (a21) at (107:2) {};
    \node[myv] (a22) at (115:2) {};
    
    \node[myv] (b11) at (35:2) {};
    \node[myv] (b12) at (43:2) {};

    \node[myv] (b21) at (73:2) {};
    \node[myv] (b22) at (65:2) {};
    
    \node[myv] (yd1) at (194:2) {};
    \node[myv] (yd2) at (202:2) {};

    \node[myv] (x11) at (251:2) {};
    \node[myv] (x12) at (259:2) {};
    
    \node[myv] (x21) at (289:2) {};
    \node[myv] (x22) at (281:2) {};

    \node[circle,dashed,red,minimum size=0.7cm,draw] (ra1) at (141:2) {};
    \node[circle,dashed,red,minimum size=0.7cm,draw] (ra2) at (111:2) {};
    \node[circle,dashed,red,minimum size=0.7cm,draw] (rb1) at (39:2) {};
    \node[circle,dashed,red,minimum size=0.7cm,draw] (rb2) at (69:2) {};
    \node[circle,dashed,red,minimum size=0.7cm,draw] (ryd) at (198:2) {};
    \node[circle,dashed,red,minimum size=0.7cm,draw] (rx1) at (255:2) {};
    \node[circle,dashed,red,minimum size=0.7cm,draw] (rx2) at (285:2) {};

    \draw (ra) -- (ra1);
    \draw (ra) -- (ra2);
    \draw (rb) -- (rb1);
    \draw (rb) -- (rb2);
    \draw (rx) -- (rx1);
    \draw (rx) -- (rx2);
    \draw (ry) -- (ryd);

    \node[myv] (a11a) at (151:2.5) {};
    \node[myv] (a12a) at (131:2.5) {};
    \node[myv] (a21a) at (101:2.5) {};
    \node[myv] (a22a) at (121:2.5) {};
    \node[myv] (b11b) at (29:2.5) {};
    \node[myv] (b12b) at (49:2.5) {};
    \node[myv] (b21b) at (79:2.5) {};
    \node[myv] (b22b) at (59:2.5) {};
    \node[myv] (x11x) at (245:2.5) {};
    \node[myv] (x12x) at (265:2.5) {};
    \node[myv] (x21x) at (295:2.5) {};
    \node[myv] (x22x) at (275:2.5) {};
    \node[myv] (yd1y) at (188:2.5) {};
    \node[myv] (yd2y) at (208:2.5) {};
    
    \draw (x11) -- (x11x);
    \draw (x12) -- (x12x);
    \draw (x21) -- (x21x);
    \draw (x22) -- (x22x);
    \draw (yd1) -- (yd1y);
    \draw (yd2) -- (yd2y);
    \draw (a11) -- (a11a);
    \draw (a12) -- (a12a);
    \draw (a21) -- (a21a);
    \draw (a22) -- (a22a);
    \draw (b11) -- (b11b);
    \draw (b12) -- (b12b);
    \draw (b21) -- (b21b);
    \draw (b22) -- (b22b);


    \end{tikzpicture}
    \caption{Construction of $G$ from $(G'=P_3,k=1,t=3,t'=1)$ by Construction~\ref{con:wbistars1}. Dashed circles 
    denote cliques. This corresponds to the reduction for \XFC{T_{3,1}}.}
    \label{fig:wbistars1}
\end{figure}

Let $H$ be $T_{t,t'}$ for $t> t'\geq 0$ and $t\geq 3$. 
Let $(G',k)$ be an instance of \DS.
We obtain $G$ from $(G',t,t',k)$ by Construction~\ref{con:wbistars1}.
An example is shown in Figure~\ref{fig:wbistars1}.

\begin{lemma}
    \label{lem:wbistars1:forward}
    Let $(G',k)$ be a yes-instance of \DS. 
    Then $(G,k)$ is a yes-instance of \XFC{H}.
\end{lemma}
\begin{proof}
    Let $D$ be a dominating set of size at most $k$ of $G'$.
    Let $F=\{wx_i| v_i\in D\}$.
    Let $w$ itself denote the vertex 
    obtained by contracting edges in $F$. 
    Since $D$ is a dominating set, $w$ is adjacent to 
    every vertex of $Y$ in $G/F$.
    We claim that $G/F$ is $H$-free.
    For a contradiction, assume that there is an induced $T_{t,t'}$
    in $G/F$
    with the central edge being $vv'$ (we do not distinguish between
    $v$ and $v'$, i.e., we do not assume that $v$ is the vertex
     attached to $t$ degree-1 vertices and $v'$ is
    the vertex attached to $t'$ degree-1 vertices in the induced $T_{t,t'}$). 
    Let $v\in A$ and $v'\in B$. Then the largest bistar graph 
    rooted at $v$ and $v'$ is $T_{t-1,t-1}$ which does not have a
    $T_{t,t'}$ as an induced subgraph. 
    The case when $v=w$ and $v'\in A\cup B$ can be proved similarly.
    Let $v\in A\cup B\cup X$ and $v'\in Y$. Then the largest bistar graph
    rooted at $v$ and $v'$ is $T_{t-1,2}$ which does not have
    an induced $T_{t,t'}$.
    Let $v=w$ and $v'\in Y$. Then the largest bistar with the 
    central edge $vv'$ is $T_{t-1,1}$, which does not have
    a $T_{t,t'}$ as an induced subgraph.
    For a pair of adjacent vertices $v,v'$ in $G/F$
    let $I(v,v')$ denote the size of a maximum independent 
    set in the graph induced by $(N(v)\setminus N[v'])\cup (N(v')\setminus N[v])$ in $G/F$.
    It is sufficient to prove that there exists no adjacent pair $(v,v')$ such that $I(v,v')\geq t+t'$. 
    Clearly,
    none of the degree-1 vertex can be in the pair we are looking for.
    The rest of the cases 
    are handled below, each of them is straight-forward to verify.
    \begin{itemize}
        \item Let $v,v'\in A'$ or $v,v'\in B'$ or $v,v'\in X'$ or
        $v,v'\in Y'$. Then $I(v,v') = t'+t'< t+t'$.
        \item Let $v,v'\in A$ or $v,v'\in B$. Then $I(v,v')=0<t+t'$.
        \item Let $v,v'\in X$ or $v,v'\in Y$. Then $I(v,v')\leq 1<t+t'$.
        \item Let $v=w$ and $v'\in X$. Then $I(v,v')\leq 1<t+t'$.
        \item Let $v=w$ and $v'\in X'$. Then $I(v,v')=t'+t-2+1=t+t'-1<t+t'$.
        \item Let $v\in A$ and $v'\in A'$, or $v\in B$ and $v'\in B'$,
        $v\in X$ and $v'\in X'$. Then $I(v',v')=t'+t-2+1=t+t'-1<t+t'$.
        \item Let $v\in Y$ and $v'\in Y'$. Then $I(v,v') = t'+2< t+t'$.
    \end{itemize}
\end{proof}

\begin{figure}
    \centering
\begin{tikzpicture}[myv/.style={circle, draw, inner sep=1.5pt}]

    \node[myv] (w) at (0,0) {};
    \node[myv] (e2) at (90:1.5) {};
    \node[myv] (e1) at (78:1.5) {};
    \node[myv] (e3) at (102:1.5) {};
    \node[myv] (a2) at (18:1.5) {};
    \node[myv] (a1) at (6:1.5) {};
    \node[myv] (a3) at (30:1.5) {};
    \node[myv] (b2) at (-54:1.5) {};
    \node[myv] (b1) at (-66:1.5) {};
    \node[myv] (b3) at (-40:1.5) {};
    \node[myv] (c2) at (234:1.5) {};
    \node[myv] (c1) at (222:1.5) {};
    \node[myv] (c3) at (246:1.5) {};
    \node[myv] (d2) at (162:1.5) {};
    \node[myv] (d1) at (150:1.5) {};
    \node[myv] (d3) at (174:1.5) {};
  
    \node[circle,dashed,red,minimum size=1.2cm,draw] (e) at (90:1.5) {};    
    \node[circle,dashed,red,minimum size=1.2cm,draw] (a) at (18:1.5) {};    
    \node[circle,dashed,red,minimum size=1.2cm,draw] (b) at (-54:1.5) {};    
    \node[circle,dashed,red,minimum size=1.2cm,draw] (c) at (234:1.5) {};    
    \node[circle,dashed,red,minimum size=1.2cm,draw] (d) at (162:1.5) {};    
    \node[myv] (e11) at (65:2.5) {};
    \node[myv] (e12) at (75:2.5) {};
    \node[myv] (e21) at (85:2.5) {};
    \node[myv] (e22) at (95:2.5) {};
    \node[myv] (e31) at (105:2.5) {};
    \node[myv] (e32) at (115:2.5) {};

    \draw (w) -- (a);
    \draw (w) -- (b);
    \draw (w) -- (c);
    \draw (w) -- (d);
    \draw (w) -- (e);
    \draw (e1) -- (e11);
    \draw (e1) -- (e12);
    \draw (e2) -- (e21);
    \draw (e2) -- (e22);
    \draw (e3) -- (e31);
    \draw (e3) -- (e32);

\end{tikzpicture}
    \caption{$K_{1,4,3,2}$ - dashed circles denote cliques}
    \label{fig:k1432}
\end{figure}
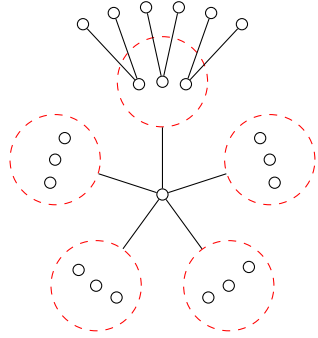

By extending the definition of $K_{1,a,b}$,
we define $K_{1,a,b,c}$ as follows.
Create a vertex $w$ and make it adjacent to $a+1$ copies of $K_b$s.
Attach $c$ degree-1 vertices to each of the $b$ vertices of one
of such $K_b$s. 
Let $w$ be known as the root of $K_{1,a,b,c}$.
An example is shown in Figure~\ref{fig:k1432}.
The following observation is the bistar counterpart of 
Observation~\ref{obs:star}.

\begin{observation}
    \label{obs:bistar}
    Let $K$ denote a copy of $K_{1,t,k+1,t'}$.
    Let $G$ has $K$ as an induced 
    subgraph such that the degree-1 vertices in the $K$ are
    degree-1 vertices in $G$.
    Then $(G,k)$ is a no-instance of \XFC{T_{t,t'}}.
\end{observation}

\begin{observation}
\label{obs:wbistars}
Let $F$ be a subset of edges of $G$ such that $|F|\leq k$ and
$G/F$ is $H$-free. Then $F$ does not contain any edge from $E[w,A]\cup E[w,B]\cup E[X,Y]$.
\end{observation}
\begin{proof}
    Assume that $wv\in F$ for some vertex $v\in A$.
    Then there is a $K_{1,t,k+1,t'}$ in $G/\{wv\}$ induced by $w, A', B, X$, and the degree-1 vertices 
    attached to the vertices of any one of the $t-1$ cliques formed by $A'$. Further, it satisfies the condition in Observation~\ref{obs:wbistars}. Then we get a contradiction
    by Observation~\ref{obs:bistar}.
    By symmetry, we obtain that $F$ does not 
    contain any edge from $E[w,B]$.
Assume that
$xy\in F$, where $x\in X, y\in Y$. 
    Then there is a $K_{1,t,k+1,t'}$ induced by $x, X', Y', A$, and the degree-1 vertices 
    attached to the vertices of $Y'$. Further, it 
    satisfies the condition in Observation~\ref{obs:wbistars}. This results 
    in a contradiction.
\end{proof}

\begin{observation}
    \label{obs:wbistars:2}
    Let $\mathcal{P}$ be a partition of vertices of $G$
    such that $G/\mathcal{P}$ is $H$-free. Further assume that
    every set in $\mathcal{P}$ is either a singleton set, or a 
    subset of $X\cup \{w\}$ or a subset of $Y$. 
    Let $P_w$ denote the set in $\mathcal{P}$ containing $w$.
    Then,
    for every set $P_y\in \mathcal{P}$ such that $P_y\subseteq Y$,
    at least one vertex of $P_w\cap X$ is adjacent to at least one
    vertex of $P_y$.
\end{observation}
\begin{proof}
    For a contradiction, assume that none of the vertices in $P_y$
    is adjacent to none of the vertices of $P_w\cap X$. Then 
    there is a $T_{t,t'}$ induced by $P_w, P_y$ and singleton sets
    corresponding to a vertex in $A$ and $t-1$ vertices (one from 
    each clique) corresponding to $A'$, and $t'$ vertices attached to one selected vertex from $A'$. Therefore, 
    $G/\mathcal{P}$ is not $T_{t,t'}$-free, which is a contradiction.
\end{proof}

\begin{lemma}
    \label{lem:wbistars1:backward}
    Let $(G,k)$ be a yes-instance of \XFC{H}.
    Then $(G',k)$ is a yes-instance of \DS.
\end{lemma}
\begin{proof}
    Let $F$ be a minimal subset of edges of $G$ such that $G/F$
    is $H$-free and $|F|\leq k$.
    By Observation~\ref{obs:wbistars}, the edges from $E[w,A]\cup E[w,B]\cup E[X,Y]$
    cannot be in $F$. It can be easily verified that, due to the 
    limited budget and the minimality of $F$, $F$ does not contain any edges of $G$ other than
    those from $E[w,X]\cup E[X,X]\cup E[Y,Y]$.
    Let $\mathcal{P}$ be the partition of vertices of $G$ corresponding
    to $F$. Let $P_w$ denote the set containing $w$.
    We create a dominating set $D$ of size at most $k$ of $G'$ as
    follows.
    Include the vertices corresponding to $P_w\cap X$ in $D$, i.e., 
    include $\{v_i|x_i\in P_w\cap X\}$ in $D$. By Observation~\ref{obs:wbistars:2}, for each $P_y\in \mathcal{P}$ such that $P_y\subseteq Y$, at least one vertex of $P_y$ is 
    adjacent to at least one vertex of $P_w\cap X$.
    Let $y\in P_y$ is adjacent to a vertex in $P_w\cap X$.
    Then for every vertex $y_i\in P_y\setminus \{y\}$,
    include $v_i$ in $D$. We claim that $D$ is a dominating set
    in $G'$. For a contradiction, assume that there is a vertex 
    $v_j$ not adjacent to any vertex in $D$. 
    Let $P_{y_j}\in \mathcal{P}$ be the set containing $y_j$.
    By the construction of $D$, for every vertex in $P_{y_j}$
    either $v_j$ or a vertex adjacent to $v_j$ is included in $D$.
    This gives us a contradiction. Now, it is sufficient to
    prove that \SURP$\leq k$. This follows from the fact that
    for $P_w$ we included $|P_w|-1$ vertices in $D$ and 
    for every set $P_y\in \mathcal{P}$ such that $P_y\subseteq Y$,
    we included at most $|P_y|-1$ vertices in $D$.
\end{proof}

Lemmas~\ref{lem:wbistars1:forward} and \ref{lem:wbistars1:backward}  
 imply that
there is a parameterized reduction from \DS\ to
\XFC{T_{t,t'}}.
\begin{lemma}
\label{lem:wbistars1}
Let $t>t'\geq 0$ and $t\geq 3$. Then 
\XFC{T_{t,t'}} is W[2]-hard.
\end{lemma}

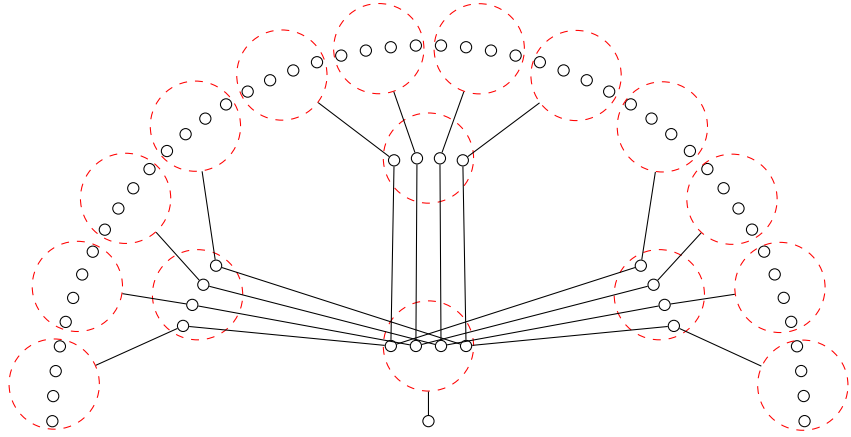
\begin{figure}
    \centering
\begin{tikzpicture}[myv/.style={circle, draw, inner sep=1.5pt}]

    \node[myv] (w) at (0,0) {};
    \foreach \x in {0,1,2,...,47}
        \node[myv] at (\x*3.83:5) {};
    \foreach \x in {0,1,2,...,11}
        \node[circle,dashed,red,minimum size=1.2cm,draw] (c\x) at (\x*15.35+5.49:5) {};
    \node[myv] (l21) at (28.725-7.5:3.5) {};    
    \node[myv] (l22) at (28.725-2.5:3.5) {};    
    \node[myv] (l23) at (28.725+2.5:3.5) {};    
    \node[myv] (l24) at (28.725+7.5:3.5) {};    
    \node[myv] (l25) at (90-7.5:3.5) {};    
    \node[myv] (l26) at (90-2.5:3.5) {};    
    \node[myv] (l27) at (90+2.5:3.5) {};    
    \node[myv] (l28) at (90+7.5:3.5) {};    
    \node[myv] (l29) at (151.275-7.5:3.5) {};    
    \node[myv] (l210) at (151.275-2.5:3.5) {};    
    \node[myv] (l211) at (151.275+2.5:3.5) {};    
    \node[myv] (l212) at (151.275+7.5:3.5) {};    
    \node[circle,dashed,red,minimum size=1.2cm,draw] (d1) at (7.5*3.83:3.5) {};    
    \node[circle,dashed,red,minimum size=1.2cm,draw] (d2) at (23.5*3.83:3.5) {};    
    \node[circle,dashed,red,minimum size=1.2cm,draw] (d3) at (39.5*3.83:3.5) {};

    \node[myv] (l11) at (0.5,1) {};
    \node[myv] (l12) at (0.167,1) {};
    \node[myv] (l13) at (-0.167,1) {};
    \node[myv] (l14) at (-0.5,1) {};

    \node[circle,dashed,red,minimum size=1.2cm,draw] (e1) at (90:1) {};

    \draw (w) -- (e1);
    \draw (l11) -- (l21);
    \draw (l11) -- (l25);
    \draw (l11) -- (l29);
    \draw (l12) -- (l22);
    \draw (l12) -- (l26);
    \draw (l12) -- (l210);
    \draw (l13) -- (l23);
    \draw (l13) -- (l27);
    \draw (l13) -- (l211);
    \draw (l14) -- (l24);
    \draw (l14) -- (l28);
    \draw (l14) -- (l212);

    \draw (l21) -- (c0);
    \draw (l22) -- (c1);
    \draw (l23) -- (c2);
    \draw (l24) -- (c3);
    \draw (l25) -- (c4);
    \draw (l26) -- (c5);
    \draw (l27) -- (c6);
    \draw (l28) -- (c7);
    \draw (l29) -- (c8);
    \draw (l210) -- (c9);
    \draw (l211) -- (c10);
    \draw (l212) -- (c11);

\end{tikzpicture}
    \caption{A $(3,3)$-canopy - each dashed circle is a clique}
    \label{fig:canopy}
\end{figure}

Now, we are left with the bistars $T_{t,t'}$ where $t=t'$.
For this, we come up with a reduction that handles more than this case. The reduction is a parameterized reduction but not a polynomial-time reduction, as the main structure, that we define next, used in the reduction has size exponential in $k$. 
For integers $t,k\geq 1$, a $(t,k)$-canopy is a graph obtained as follows. The vertices are arranged in 
$k+1$ levels - level 0 to level $k$. 
The set of vertices in $i$\textsuperscript{th} level is 
denoted by $L_i$.
The set $L_0$ contains a single vertex, which is called the root.
For $1\leq i\leq k$, $L_i$ contains $t^{\lfloor i/2\rfloor}(k+1)^{\lfloor(i-1)/2\rfloor}$ cliques, 
denoted by $L_{i,1}, L_{i,2},\ldots$, 
each
of size $k+1$. 
The single vertex in $L_0$ is adjacent to all vertices in the 
clique $L_1$.
For $1\leq i\leq k-1$,  
the edges between cliques in $L_i$ and cliques in $L_{i+1}$
are as follows. 
If $i$ is odd, then for every
clique $L_{i,j}$ there are $t$ unique cliques in $L_{i+1}$ such 
that edges of a perfect matching is added between $L_{i,j}$
and each such clique, i.e., for every vertex in $L_{i,j}$, there
is a unique neighbor in each such clique .
If $i$ is even, then every vertex in a clique $L_{i,j}$
is adjacent to all vertices of a 
unique clique in $L_{i+1}$. An example is shown in Figure~\ref{fig:canopy}.

\begin{construction}
    \label{con:wbistars:2}
    Let $G'$ be a graph with $n$ vertices $\{v_1,v_2,\ldots,v_n\}$ and let $k,t,t'$ be 
    integers. A graph $G$ is constructed from $(G',k,t,t')$
    as follows.
    \begin{itemize}
        \item Introduce a set of $t$ cliques each of size $k+1$. 
        These cliques are denoted by $A_1,A_2,\ldots,A_t$, and 
        the set of all these vertices is denoted by $A$.
        \item Introduce a clique $X=\{x_1,x_2,\ldots,x_n\}$ of $n$ vertices. 
        \item Introduce a vertex $w$ and make it adjacent to $A\cup X$.
        \item Introduce $t'$ cliques each of size $n$. These
        cliques are denoted by $Y_1,Y_2,\ldots,Y_{t'}$ and the set of 
        all these vertices is denoted by $Y$.
        Let the vertices in $Y_1$ be $\{y_{1,1}, y_{1,2},\ldots,y_{1,n}\}$.
        \item The vertices $x_i$ is adjacent to $y_{1,j}$
        if and only if $i=j$ or $v_i$ and $v_j$ are adjacent in $G'$.
        \item For $i=2$ to $t'$, the edges of a perfect matching
        is added between $X$ and $Y_{i}$, i.e., each vertex in $X$
        is made adjacent to a unique vertex in $Y_i$.
        \item For each vertex in $y\in Y$, introduce a $(t',k)$-canopy and identify $y$ with the root of the canopy.
    \end{itemize}
\end{construction}

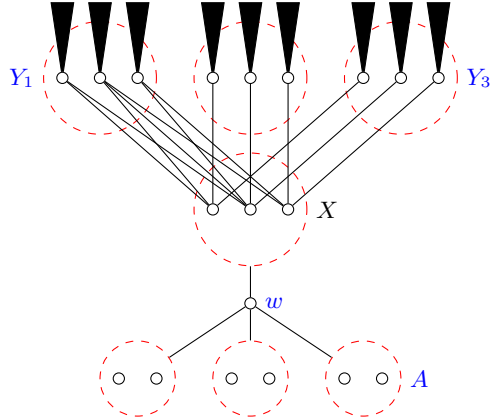
\begin{figure}
    \centering
\begin{tikzpicture}[myv/.style={circle, draw, inner sep=1.5pt}]

    \node[myv,label={[blue]0:$w$}] (w) at (0,0) {};
    \node[myv] (a11) at (-1.75,-1) {};
    \node[myv] (a12) at (-1.25,-1) {};
    \node[myv] (a21) at (-0.25,-1) {};
    \node[myv] (a22) at (0.25,-1) {};
    \node[myv] (a31) at (1.25,-1) {};
    \node[myv] (a32) at (1.75,-1) {};
    \node[circle,dashed,red,minimum size=1.0cm,draw] (a2) at (0,-1) {};
    \node[circle,dashed,red,minimum size=1.0cm,draw] (a1) at (-1.5,-1) {};
    \node[circle,dashed,red,minimum size=1.0cm,draw,label={[blue]0:$A$}] (a3) at (1.5,-1) {};    
    \draw (w) -- (a1);
    \draw (w) -- (a2);
    \draw (w) -- (a3);

    \node[myv] (x2) at (0,1.25) {};
    \node[myv] (x1) at (-0.5,1.25) {};
    \node[myv] (x3) at (0.5,1.25) {};
    \node[circle,dashed,red,minimum size=1.5cm,draw,label=0:$X$] (x) at (0,1.25) {};  
    \draw (w) -- (x);

    \node[myv] (y22) at (0,3) {};
    \node[myv] (y21) at (-0.5,3) {};
    \node[myv] (y23) at (0.5,3) {};
    \node[circle,dashed,red,minimum size=1.5cm,draw] (y2) at (0,3) {};  

    \node[myv] (y12) at (-2,3) {};
    \node[myv] (y11) at (-2.5,3) {};
    \node[myv] (y13) at (-1.5,3) {};
    \node[circle,dashed,red,minimum size=1.5cm,draw,label={[blue]180:$Y_1$}] (y1) at (-2,3) {};  

    \node[myv] (y32) at (2,3) {};
    \node[myv] (y31) at (1.5,3) {};
    \node[myv] (y33) at (2.5,3) {};
    \node[circle,dashed,red,minimum size=1.5cm,draw,label={[blue]0:$Y_3$}] (y3) at (2,3) {};  

    \draw (x1) -- (y11);
    \draw (x1) -- (y12);
    \draw (x2) -- (y11);
    \draw (x2) -- (y12);
    \draw (x2) -- (y13);
    \draw (x3) -- (y12);
    \draw (x3) -- (y13);

    \draw (x1) -- (y21);
    \draw (x1) -- (y31);
    \draw (x2) -- (y22);
    \draw (x2) -- (y32);
    \draw (x3) -- (y23);
    \draw (x3) -- (y33);

    \draw[fill=black] (y12) -- (-2.15,4) -- (-1.85,4) -- (y12);
    \draw[fill=black] (y11) -- (-2.65,4) -- (-2.35,4) -- (y11);
    \draw[fill=black] (y13) -- (-1.65,4) -- (-1.35,4) -- (y13);
    \draw[fill=black] (y21) -- (-0.65,4) -- (-0.35,4) -- (y21);
    \draw[fill=black] (y22) -- (-0.15,4) -- (0.15,4) -- (y22);
    \draw[fill=black] (y23) -- (0.35,4) -- (0.65,4) -- (y23);
    \draw[fill=black] (y31) -- (1.35,4) -- (1.65,4) -- (y31);
    \draw[fill=black] (y32) -- (1.85,4) -- (2.15,4) -- (y32);
    \draw[fill=black] (y33) -- (2.35,4) -- (2.65,4) -- (y33);

\end{tikzpicture}
    \caption{Construction of $G$ from $(G'=P_3,k=1,t=3,t'=3)$ by
    Construction~\ref{con:wbistars:2}. Dashed circles denote cliques
    and dark triangles denote copies of $(t',k)$-canopy.} 
    \label{fig:wbistars2}
\end{figure}

An example of the construction is given in Figure~\ref{fig:wbistars2}.
Let $(G',k)$ be an instance of \DS. 
Let $H$ be $T_{t,t'}$, where $t\geq 3$ and $3\leq t'\leq t$.
Let $G$ be constructed from $(G',k,t,t')$ by Construction~\ref{con:wbistars:2}.

\begin{lemma}
    \label{lem:wbistars2:forward}
    Let $(G',k)$ be a yes-instance of \DS.
    Then $(G,k)$ is a yes-instance of \XFC{H}.
\end{lemma}
\begin{proof}
    Let $D$ be a dominating set of size at most $k$ of $G'$.
    Let $F=\{wx_i|v_i\in D\}$. Clearly $|F|=|D|\leq k$.
    We claim that $G/F$ is $H$-free. Let $w$ itself
    denote the vertex obtained by contracting the edges in $F$.
    Since $D$ is a dominating set of $G'$, $w$ is adjacent 
    to all vertices of $Y_{1}$ in $G/F$.
    For a contradiction assume that there is an induced 
    $T_{t,t'}$ in $G/F$ with the central edge $vv'$.  
    If $v=w$ and $v'\in X$, then the largest bistar where $vv'$
    is the central edge is $T_{t+1,t'-1}$ which does not have
    an induced $T_{t,t'}$. 
    If $v=w$ and $v'\in Y$, then the largest bistar where $vv'$
    is the central edge is $T_{t+t',1}$ which does not 
    have an induced $T_{t,t'}$.
    For a pair of adjacent vertices $v,v'$ in $G/F$
    let $I(v,v')$ denote the size of a maximum independent 
    set in the graph induced by $(N(v)\setminus N[v'])\cup (N(v')\setminus N[v])$ in $G/F$.
    It is sufficient to prove that there exists no adjacent pair $(v,v')$ such that $I(v,v')\geq t+t'$. 
    The remaining cases are handled below, 
    each of them is easy to verify. 
    \begin{itemize}
        \item Let $v,v'\in K$ where $K$ is a clique 
        in an odd-numbered level in any $(t',k)$-canopy 
        attached to the vertices of $Y$. Then $I(v,v')\leq t'<t+t'$.
        \item Let $v,v'\in K$ where $K$ is a clique 
        in an $i$\textsuperscript{th} level, for any even $i\geq 2$ in any $(t',k)$-canopy 
        attached to the vertices of $Y$. Then $I(v,v')= 3<t+t'$.
        \item Let $v,v'\in Y_i$, for $1\leq i\leq t'$.
        Then $I(v,v')=3<t+t'$.
        \item Let $v,v'\in X$. Then $I(v,v')=t'<t+t'$.
        \item Let $v,v'\in A$. Then $I(v,v')=0$.
        \item Let $v\in L_i$ and $v'\in L_{i+1}$, where $0\leq i\leq k-1$ for a $(t',k)$-canopy attached to a vertex in $Y$.
        Then $I(v,v')\leq t'+2\leq t+t'$.
        \item Let $v\in X$ and $v'\in Y$. 
        Then $I(v,v')\leq 2+t'<t+t'$.
        \item Let $v=w, v'\in A$. Then $I(v,v')=0$.
    \end{itemize}
\end{proof}

The following observation is straight-forward to verify.
\begin{observation}
    \label{obs:wbistars2:0}
    Let $\mathcal{P}$ be a partition of vertices of $G$
    such that each set induces a connected graph and $G/\mathcal{P}$
    is $H$-free and \SURP$\leq k$. 
    Then for every clique $K$ introduced while constructing $G$, 
    there is a subset $K'$ of $K$ such that $K'\in \mathcal{P}$.
\end{observation}

\begin{observation}
    \label{obs:wbistars2}
    Let $\mathcal{P}$ be a partition of vertices of $G$
    such that each set induces a connected graph, $G/\mathcal{P}$
    is $H$-free, and \SURP$\leq k$. Further assume that $\mathcal{P}$
    is such a partition with least \SURP. Let $P_w\in \mathcal{P}$
    be the set containing $w$. Then $P_w$ does not contain 
    any vertex in any of the $(t,k)$-canopies attached to vertices
    in $Y$.
\end{observation}
\begin{proof}
    For a contradiction, assume that $P_w$
    contains a $(t',k)$-canopy vertex and let $i$ be the largest
    index such that $P_w$ contains a vertex in $L_i$
    among all $(t,k)$-canopies attached to vertices of $Y$.
    Since the budget is only $k$, we obtain that $i\leq k-2$.
    Let $v$ be any vertex
    in $P_w\cap L_i$. 
    Assume that $i$ is an even number. 
    Then there is a $T_{t,t'}$ induced by $P_w$, $t$ sets
    in $\mathcal{P}$ subsets of $t$ cliques in $A$ (Observation~\ref{obs:wbistars2:0}), a set $P_{v'}$ corresponding to a clique in $L_{i+1}$ adjacent to
    $v$, and $t'$ sets corresponding to the $t'$ vertices in $L_{i+2}$ adjacent 
    to $P_{v'}$.
    Let $i\geq 1$ be an odd number.
    Then there is a $T_{t,t'}$ induced by $P_w$, $t$ sets
    in $\mathcal{P}$ subsets of $t$ cliques in $A$ (Observation~\ref{obs:wbistars2:0}), a set $P_{v'}$ corresponding to the clique in $v$ is part of (such that $v\notin P_{v'}$), 
    and $t'$ sets corresponding to the vertices in $L_{i+1}$
    adjacent to $P_{v'}$.
\end{proof}
\begin{lemma}
    \label{lem:wbistars2:backward}
    Let $(G,k)$ be a yes-instance of \XFC{H}.
    Then $(G',k)$ is a yes-instance of \DS.
\end{lemma}
\begin{proof}
    Let $\mathcal{P}$ be a partition of vertices of $G$
    into connected subgraphs such that $G/\mathcal{P}$ 
    is $H$-free and \SURP$\leq k$. 
    Further assume that $\mathcal{P}$ is having the least 
    \SURP.
    Let 
    $P_w\in \mathcal{P}$ denote the set containing $w$.
    By Observation~\ref{obs:wbistars:2}, 
    $P_w$ does not contain any vertex in the $(t,k)$-canopies 
    introduced during the construction of $G$.
    Now, as we have done in the proof of Lemma~\ref{lem:wbistars1:backward}, we can construct a 
    dominating set $D$ of $G'$: Add the vertices in $G'$ corresponding to $P_w\cap X$ in $D$.
    For every $P_y\in \mathcal{P}$ such that $y\in Y_1$, 
    we know that at least one vertex $v$ of $P_y$ must be 
    adjacent to $P_w\cap X$. For all other vertices $v_i$ in $P_y\cap Y_1$,
    add $v_i$ in $D$. It is straight-forward to verify that $D$
    is a dominating set of size at most $k$ of $G'$.
\end{proof}

Lemmas~\ref{lem:wbistars2:forward} and \ref{lem:wbistars2:backward}  
imply that there is a parameterized reduction from \DS\ to
\XFC{T_{t,t'}}.

\begin{lemma}
\label{lem:wbistars2}
Let $t\geq t'\geq 3$.
Then \XFC{T_{t,t'}} is W[2]-hard.
\end{lemma}

Now, Lemmas~\ref{lem:wgeneral}, \ref{lem:wbistars1}, and \ref{lem:wbistars2} imply the main result of this section.
\begin{theorem}
    \label{thm:whard}
    Let $T$ be a tree which is neither a star of at most 4 vertices ($\{K_1, K_2,P_3, K_{1,3}\}$) nor a bistar in $\{T_{1,1}, T_{2,1}, T_{2,2}\}$. Then \XFC{T_{t,t'}} is W[2]-hard.
\end{theorem}

We believe that our W[2]-hardness result on trees will be a stepping stone
for an eventual parameterized complexity classification of \XFC{H}.
The most challenging hurdle for such a complete classification can be the graphs $H$ where each component is of at most 2 vertices, and the case of claw, the usual trouble-maker for other graph modification problems to $H$-free graphs.

We conclude with some folklore observations. As noted in a version of \cite{DBLP:conf/iwpec/LokshtanovMS13}, the property that ``there exists at most $k$ edges contracting which
results in an $H$-free graph'' can be expressed in \MSOO. The length of the corresponding \MSOO\ formula will be a function of $k$. Then, there exists FPT algorithms for \XFC{H}, whenever $H$-free graphs have bounded rankwidth (See Chapter 7 of the textbook \cite{DBLP:books/sp/CyganFKLMPPS15}). 
This, in particular, implies that \XFC{K_2+K_1} can be solved in FPT time. 
It is known that every component of a paw-free graph is either triangle-free or complete multipartite~\cite{DBLP:journals/ipl/Olariu88}, where
where paw is the graph \begin{tikzpicture}[myv/.style={circle, draw, inner sep=1.5pt}]
    \node[myv] (o) at (0,0) {};
    \node[myv] (v1) at (-0.3,0.15) {};
    \node[myv] (v2) at (-0.3,-0.15) {};
    \node[myv] (v3) at (0.3,0) {};

    \draw (o) -- (v1);
    \draw (o) -- (v2);
    \draw (v1) -- (v2);
    \draw (o) -- (v3);
\end{tikzpicture}
. Then the existance of FPT algorithms for \XFC{K_3} and \XFC{K_2+K_1} imply that there exists an FPT algorithm for \XFC{paw}.

 \bibliographystyle{splncs04}
 \bibliography{main}

\begin{thebibliography}{10}
\providecommand{\url}[1]{\texttt{#1}}
\providecommand{\urlprefix}{URL }
\providecommand{\doi}[1]{https://doi.org/#1}

\bibitem{AravindSS17}
Aravind, N.R., Sandeep, R.B., Sivadasan, N.: Dichotomy results on the hardness
  of {H}-free edge modification problems. {SIAM} J. Discrete Math.
  \textbf{31}(1),  542--561 (2017). \doi{10.1137/16M1055797},
  \url{https://doi.org/10.1137/16M1055797}

\bibitem{asano1983edge}
Asano, T., Hirata, T.: Edge-contraction problems. Journal of Computer and
  System Sciences  \textbf{26}(2),  197--208 (1983)

\bibitem{DBLP:journals/dam/BelmonteHH12}
Belmonte, R., Heggernes, P., van~'t Hof, P.: Edge contractions in subclasses of
  chordal graphs. Discret. Appl. Math.  \textbf{160}(7-8),  999--1010 (2012).
  \doi{10.1016/j.dam.2011.12.012},
  \url{https://doi.org/10.1016/j.dam.2011.12.012}

\bibitem{DBLP:journals/dam/Bonuccelli85}
Bonuccelli, M.A.: Dominating sets and domatic number of circular arc graphs.
  Discret. Appl. Math.  \textbf{12}(3),  203--213 (1985).
  \doi{10.1016/0166-218X(85)90025-3},
  \url{https://doi.org/10.1016/0166-218X(85)90025-3}

\bibitem{brouwer1987contractibility}
Brouwer, A.E., Veldman, H.J.: Contractibility and np-completeness. Journal of
  Graph Theory  \textbf{11}(1),  71--79 (1987)

\bibitem{DBLP:journals/ipl/Cai96}
Cai, L.: Fixed-parameter tractability of graph modification problems for
  hereditary properties. Inf. Process. Lett.  \textbf{58}(4),  171--176 (1996).
  \doi{10.1016/0020-0190(96)00050-6},
  \url{https://doi.org/10.1016/0020-0190(96)00050-6}

\bibitem{CaiC15incompressibility}
Cai, L., Cai, Y.: Incompressibility of {H}-free edge modification problems.
  Algorithmica  \textbf{71}(3),  731--757 (2015).
  \doi{10.1007/s00453-014-9937-x}

\bibitem{DBLP:conf/iwpec/CaiG13}
Cai, L., Guo, C.: Contracting few edges to remove forbidden induced subgraphs.
  In: Gutin, G.Z., Szeider, S. (eds.) Parameterized and Exact Computation - 8th
  International Symposium, {IPEC} 2013, Sophia Antipolis, France, September
  4-6, 2013, Revised Selected Papers. Lecture Notes in Computer Science,
  vol.~8246, pp. 97--109. Springer (2013). \doi{10.1007/978-3-319-03898-8\_10},
  \url{https://doi.org/10.1007/978-3-319-03898-8\_10}

\bibitem{DBLP:books/sp/CyganFKLMPPS15}
Cygan, M., Fomin, F.V., Kowalik, L., Lokshtanov, D., Marx, D., Pilipczuk, M.,
  Pilipczuk, M., Saurabh, S.: Parameterized Algorithms. Springer (2015).
  \doi{10.1007/978-3-319-21275-3},
  \url{https://doi.org/10.1007/978-3-319-21275-3}

\bibitem{DBLP:journals/jgaa/Eppstein09}
Eppstein, D.: Finding large clique minors is hard. J. Graph Algorithms Appl.
  \textbf{13}(2),  197--204 (2009). \doi{10.7155/jgaa.00183},
  \url{https://doi.org/10.7155/jgaa.00183}

\bibitem{DBLP:books/fm/GareyJ79}
Garey, M.R., Johnson, D.S.: Computers and Intractability: {A} Guide to the
  Theory of NP-Completeness. W. H. Freeman (1979)

\bibitem{garland1997surface}
Garland, M., Heckbert, P.S.: Surface simplification using quadric error
  metrics. In: Proceedings of the 24th annual conference on Computer graphics
  and interactive techniques. pp. 209--216 (1997)

\bibitem{DBLP:journals/tcs/GolovachKPT13}
Golovach, P.A., Kaminski, M., Paulusma, D., Thilikos, D.M.: Increasing the
  minimum degree of a graph by contractions. Theor. Comput. Sci.  \textbf{481},
   74--84 (2013). \doi{10.1016/j.tcs.2013.02.030},
  \url{https://doi.org/10.1016/j.tcs.2013.02.030}

\bibitem{DBLP:journals/ipl/GuillemotM13}
Guillemot, S., Marx, D.: A faster {FPT} algorithm for bipartite contraction.
  Inf. Process. Lett.  \textbf{113}(22-24),  906--912 (2013).
  \doi{10.1016/j.ipl.2013.09.004},
  \url{https://doi.org/10.1016/j.ipl.2013.09.004}

\bibitem{DBLP:phd/ndltd/Guo13}
Guo, C.: Parameterized complexity of graph contraction problems. Ph.D. thesis,
  Chinese University of Hong Kong, Hong Kong (2013),
  \url{http://library.cuhk.edu.hk/record=b5549784}

\bibitem{DBLP:journals/siamdm/HeggernesHLP13}
Heggernes, P., van~'t Hof, P., Lokshtanov, D., Paul, C.: Obtaining a bipartite
  graph by contracting few edges. {SIAM} J. Discret. Math.  \textbf{27}(4),
  2143--2156 (2013). \doi{10.1137/130907392},
  \url{https://doi.org/10.1137/130907392}

\bibitem{DBLP:journals/dm/HoedeV81}
Hoede, C., Veldman, H.J.: Contraction theorems in hamiltonian graph theory.
  Discret. Math.  \textbf{34}(1),  61--67 (1981).
  \doi{10.1016/0012-365X(81)90022-4},
  \url{https://doi.org/10.1016/0012-365X(81)90022-4}

\bibitem{DBLP:journals/ipl/KaplanS94}
Kaplan, H., Shamir, R.: The domatic number problem on some perfect graph
  families. Inf. Process. Lett.  \textbf{49}(1),  51--56 (1994).
  \doi{10.1016/0020-0190(94)90054-X},
  \url{https://doi.org/10.1016/0020-0190(94)90054-X}

\bibitem{DBLP:journals/jcss/LewisY80}
Lewis, J.M., Yannakakis, M.: The node-deletion problem for hereditary
  properties is {NP}-{C}omplete. J. Comput. Syst. Sci.  \textbf{20}(2),
  219--230 (1980). \doi{10.1016/0022-0000(80)90060-4},
  \url{https://doi.org/10.1016/0022-0000(80)90060-4}

\bibitem{DBLP:conf/iwpec/LokshtanovMS13}
Lokshtanov, D., Misra, N., Saurabh, S.: On the hardness of eliminating small
  induced subgraphs by contracting edges. In: Gutin, G.Z., Szeider, S. (eds.)
  Parameterized and Exact Computation - 8th International Symposium, {IPEC}
  2013, Sophia Antipolis, France, September 4-6, 2013, Revised Selected Papers.
  Lecture Notes in Computer Science, vol.~8246, pp. 243--254. Springer (2013).
  \doi{10.1007/978-3-319-03898-8\_21},
  \url{https://doi.org/10.1007/978-3-319-03898-8\_21}

\bibitem{lovasz2006graph}
Lov{\'a}sz, L.: Graph minor theory. Bulletin of the American Mathematical
  Society  \textbf{43}(1),  75--86 (2006)

\bibitem{DBLP:journals/jcss/MarxS22}
Marx, D., Sandeep, R.B.: Incompressibility of {H}-free edge modification
  problems: Towards a dichotomy. J. Comput. Syst. Sci.  \textbf{125},  25--58
  (2022). \doi{10.1016/j.jcss.2021.11.001},
  \url{https://doi.org/10.1016/j.jcss.2021.11.001}

\bibitem{DBLP:journals/ipl/Olariu88}
Olariu, S.: Paw-fee graphs. Inf. Process. Lett.  \textbf{28}(1),  53--54
  (1988). \doi{10.1016/0020-0190(88)90143-3},
  \url{https://doi.org/10.1016/0020-0190(88)90143-3}

\end{thebibliography}
\end{document}